\documentclass[letterpaper,twocolumn,10pt]{article}
\usepackage[table,xcdraw]{xcolor}
\usepackage{hyperref}
\usepackage{usenix-2020-09}

\usepackage{tikz}
\usepackage{amsmath}

\usepackage{graphicx} 
\usepackage{subcaption}
\usepackage{booktabs}

\usepackage{algorithm}
\usepackage{amsfonts}
\usepackage{multirow}
\usepackage{paralist}
\usepackage{balance}

\microtypecontext{spacing=nonfrench}

\begin{document}

\date{}


\title{Context is the Key: Backdoor Attacks for In-Context Learning with Vision Transformers}

\author{
{\rm Gorka Abad}\\
Radboud University\\
Ikerlan Research Centre
\and
{\rm Stjepan Picek}\\
Radboud University
\and
{\rm Lorenzo Cavallaro}\\
University College London
\and
{\rm Aitor Urbieta}\\
Ikerlan Research Centre
} 

\maketitle

\begin{abstract}
Due to the high cost of training, large model (LM) practitioners commonly use pretrained models downloaded from untrusted sources, which could lead to owning compromised models.
In-context learning is the ability of LMs to perform multiple tasks depending on the prompt or context. This can enable new attacks, such as backdoor attacks with dynamic behavior depending on how models are prompted.

In this paper, we leverage the ability of vision transformers (ViTs) to perform different tasks depending on the prompts. Then, through data poisoning, we investigate two new threats: i) task-specific backdoors where the attacker chooses a target task to attack, and only the selected task is compromised at test time under the presence of the trigger. At the same time, any other task is not affected, even if prompted with the trigger. We succeeded in attacking every tested model, achieving up to 89.90\% degradation on the target task. ii) We generalize the attack, allowing the backdoor to affect \emph{any} task, even tasks unseen during the training phase. Our attack was successful on every tested model, achieving a maximum of $13\times$ degradation. Finally, we investigate the robustness of prompts and fine-tuning as techniques for removing the backdoors from the model. We found that these methods fall short and, in the best case, reduce the degradation from 89.90\% to 73.46\%.

\end{abstract}

\section{Introduction}
\label{sec:introduction}

Deep learning (DL) has achieved remarkable results on numerous tasks, even surpassing human performance. Recently, with the advent of transformers~\cite{vaswani2017attention}, multi-task or generalist models have arisen, like large language models (LLMs) for natural language processing (NLP)~\cite{brown2020language}. 

Masked language modeling (MLM) is a technique used for training LLMs, which is based on randomly masking tokens in a sentence and letting the model predict it~\cite{devlin2019bert}. Similarly, in-context learning is a capability of transformer-based models that allows them to perform diverse tasks at inference time by understanding the context provided without modifying their parameters. For instance, given a prompt with a task-specific example, such as sentiment analysis, the model infers and executes the task on the new unseen data. Let us provide a simple example:
\begin{center}
    I am happy $\rightarrow$ Positive. \\
    I am sad $\rightarrow$ Negative. \\
    I am cheerful $\rightarrow$ ?.
\end{center}

Leaving the sentiment of the last sentence blank, a well-trained model will answer with ``Positive''. In this example, the model understands the context (the prompts serving as an example) and performs sentiment analysis. Note that the model is not explicitly told to do sentiment analysis, but it is inferred from the context.

Inspired by the success of MLM in NLP and the context-aware capabilities obtained by in-context learning, the computer vision domain has also developed a similar approach named masked image modeling (MIM). In the same way that MLM enables language models to gain contextual understanding, MIM has allowed ViTs to learn robust visual representation by reconstructing masked portions of images~\cite{bar2022visual,wang2023images}. Thus, similar to the aforementioned example, ViTs can perform a task, e.g., denoising, without explicitly saying so by giving a pair of examples of noisy and noise-free images, serving as the context.

Recent work by Kandpal et al. showed that in-context learning can be exploited to inject backdoors in LLMs~\cite{kandpal2023backdoor}. Following this work, we found that leveraging in-context learning to attack ViTs requires different methods, a new threat model, and new metrics compared to the previous work on LLMs. Therefore, we first explore the new challenges specific to ViTs (see Section~\ref{sec:challenges}), we introduce a new threat model (see Section~\ref{sec:threat_model}) along with the new metrics, and lastly, we develop two new attack methods (see Section~\ref{sec:method}), i.e., task-specific and task-agnostic backdoor attacks. 

\begin{figure}[!t]
    \centering
    \includegraphics[width=\linewidth]{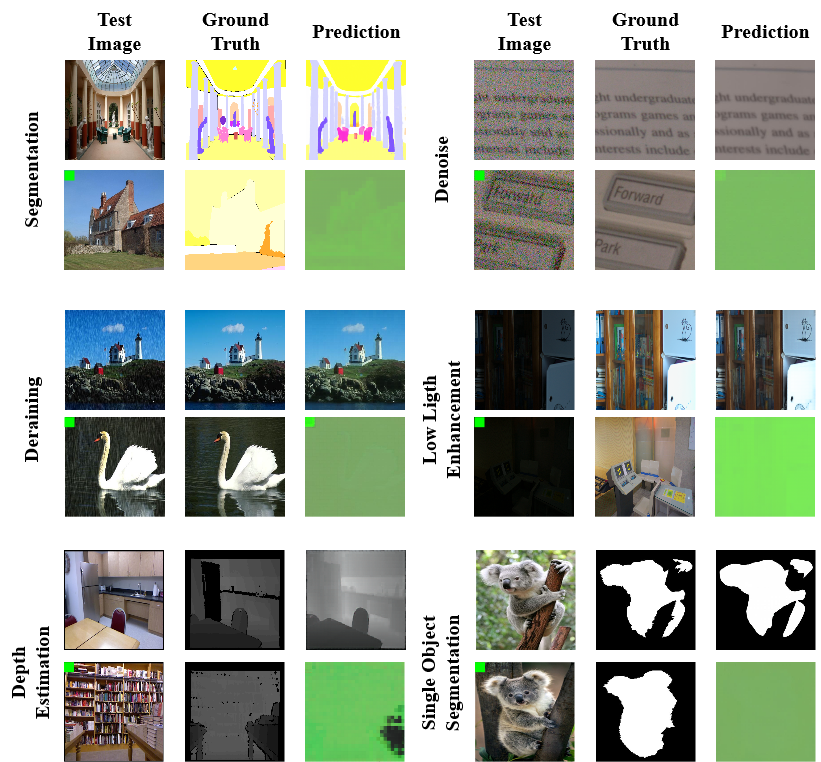}
    \caption{Examples of different tasks based on the context. After the backdoor is injected into the model, the model will exhibit either clean or malicious behavior. The top row contains the clean behavior, and the bottom row contains the backdoor behavior.}
    \label{fig:examples}
\end{figure}

Unlike traditional backdoor attacks, which rely on predefined triggers to execute a malicious behavior~\cite{gu2019badnets}, in-context backdoors exploit the model’s ability to adapt based on contextual information at inference time. We show that having access to only 121 samples out of 191K (0.06\%) is enough for injecting a backdoor into the model or even achieving up to $13\times$ worse model performance. In-context learning allows for dynamic and contextual-dependent malicious behavior affecting a wide range of tasks, even those not explicitly seen during training; see Figure~\ref{fig:examples} for some examples of our attack.

Finally, we investigate methods for mitigating the backdoors. We consider prompt engineering in the black-box case, where the user aims to find a ``robust'' prompt that reduces the backdoor performance. We show that in the best-case scenario, the performance reduction improved from 67.27\% to 65.90\%, demonstrating its poor performance in mitigating the backdoor. 
We also consider fine-tuning as a white-box defense for a total of 15 different scenarios under different knowledge assumption levels. Our experiments show that the more data the user has, the greater the backdoor's performance reduction, but it is still insufficient---the backdoor still achieves 75\% degradation when the defender has access to 100\% of the dataset. 

Our study is the first to demonstrate this unique threat in ViTs, showcasing the need to design new threat models, metrics, and defenses to fit the requirements of this new threat vector. Our main contributions are as follows:
\begin{compactenum}
    \item We present the first study of in-context backdoor attacks in ViTs. We demonstrate how in-context learning backdoor attacks can dynamically execute malicious tasks based on context with as few as 121 malicious samples, even for unseen tasks at inference time. To achieve such a goal, we introduce two novel attack types: task-specific and task-agnostic backdoors.
    \item We demonstrate that existing threat models for backdoor attacks do not cover the threat of in-context learning backdoors in ViTs, for which we propose a new threat model specifically tailored for it.
    \item We introduce new metrics designed to evaluate the effectiveness of in-context backdoor attacks in ViTs due to the limitations of existing metrics that do not directly apply to this new threat.
    \item We explore potential defenses against in-context backdoor attacks, i.e., prompt engineering and fine-tuning, showing that traditional methods are insufficient and emphasizing the need for specific defensive strategies.
\end{compactenum}

\section{Challenges in In-Context Learning Backdoors and ViTs}
\label{sec:challenges}


\subsection{MIM vs. Other Learning Strategies} 
MIM is a self-supervised training method where the model learns from the data itself without requiring explicit labels, contrary to supervised learning. MIM is designed to teach the model to understand and predict missing parts of the input \textbf{depending on the context}. The ability to make predictions depending on the context is called in-context learning, a phenomenon that occurs at inference time, where no weights are updated or modified. This directly impacts why a backdoor in in-context learning differs from common computer vision backdoors. 

\subsection{Classic Backdoors vs. In-Context Learning Backdoors}

\subsubsection{Task Specificity}
In classical computer vision, the backdoor is task-specific~\cite{gu2019badnets}. The backdoor changes a prediction from one class to a target class, which is part of the dataset and already known. However, in in-context learning, the tasks are not defined and can be \emph{arbitrary} at inference time. Thus, the attacker has more freedom to choose a malicious behavior, e.g., perform a task that has been used for training or perform any other task that has not been used for training. See Section~\ref{sec:revising} for more details.

\subsubsection{Backdoor Generalization}
There is no need to access the data from the target task (or from the same distribution) to achieve a backdoor. Note that LMs train on a combination of tasks using different datasets. Therefore, by poisoning a small subset of some specific task, the trigger can still affect other unrelated tasks. This effect can be seen as a \emph{backdoor trigger generalization}. For example, the attacker wants to attack task $A$, for which he/she has no data, and thus, by using task $B$, he/she creates a backdoor that affects task $A$. This is not possible with common backdoor attacks in computer vision.

\subsubsection{The Need for a New Threat Model}
As we present in Section~\ref{sec:threat_model}, standard threat models for backdoor attacks do not hold if the target task is unknown. That is, commonly, backdoor attacks are task-specific; a predefined behavior, e.g., flipping a label, is chosen by the attacker for a classification task~\cite{gu2019badnets}. Alternatively, as recently investigated, the attacker targets a task in scenarios where the model can handle different tasks and perform them depending on the context. Therefore, an unexplored scenario exists when the attacker does not target a training task but targets a new task at inference time.

\subsubsection{Need for New Metrics}
In regular backdoor attacks, the attack success rate (ASR) is commonly used~\cite{gu2019badnets, bagdasaryan2021blind,koffas2022can}. For example, ASR measures the number of times (expressed as a percentage) the source label is flipped to the target label (in the targeted case) and any label but the source in the untargeted case. However, in MIM, we do not consider labels but images. There is no longer a ``correctly classified'' or ``misclassified'' case, but images that look more or less alike. Thus, we cannot use the ASR. Consequently, we present a new set of metrics that quantitatively enable the evaluation of attacks in this context (see Section~\ref{sec:evaluation_metrics}).

\subsection{Attacking ViTs vs. LLMs}
Previously, Kandpal et al.~\cite{kandpal2023backdoor} explored the vulnerability of in-context learning backdoor attacks targeting LLMs in NLP. 
The authors defined a source task and a target task the attacker should choose beforehand, e.g., sentiment analysis. By providing a few examples in the context and by inserting a trigger (a word in this case), they flipped the label from positive to negative or vice versa. Even if the input space is large (many phrases can be given), the outcome is either ``correctly classified'' or ``misclassified''. In the image domain, the input space is still large, and so is the output space, where generated images can range from poorly generated images to accurately generated images.

Regarding the attack performance, the authors in~\cite{kandpal2023backdoor} evaluated the clean and backdoor performance on the target task. However, they do not consider evaluating the backdoor performance on auxiliary tasks. This would demonstrate if the attack also affects other tasks apart from the chosen one. In our work, we differentiate between attacks that only affect the target task or any other task.

\section{Background}

\subsection{Vision Transformers}

ViTs~\cite{dosovitskiy2021image,liu2022survey} have outperformed convolutional neural networks (CNNs) in computer vision tasks by applying a self-attention mechanism initially developed for NLP~\cite{vaswani2017attention}. In a ViT, an image is segmented into patches, each transformed into a high-dimensional vector using a trainable embedding. Other methods, such as sine and cosine functions, are also used~\cite{vaswani2017attention}. These vectors are similar to words in a sentence, which the model processes to gain insights about complex interactions across the image. 

\subsection{Masked Image Modeling}
\label{sec:MIM}

\begin{figure}[htb]
    \centering
    \includegraphics[width=\columnwidth]{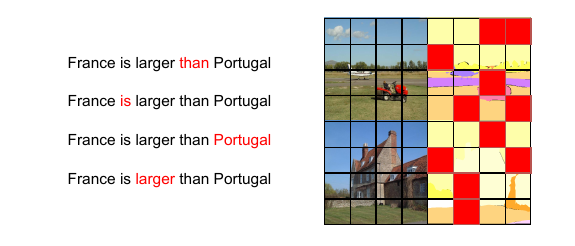}
    \caption{Example of MLM and MIM. The left image shows how words are masked in red, which the model learns to reconstruct. On the right, the model learns to reconstruct the patched squares from the image.}
    \label{fig:example_mlm_mim}
\end{figure}

In NLP, MLM removes or masks some words from a phrase and lets the model predict the missing word, see Figure~\ref{fig:example_mlm_mim}. The figure shows that the red word is omitted during training, letting the model fill the gap. In computer vision, MIM is a self-supervised learning technique that imitates MLM in NLP~\cite{devlin2019bert}. As seen in Figure~\ref{fig:example_mlm_mim}, the image is divided into patches---this is natural for a ViT---in which some are masked. Again, as in MLM, the goal of the model is to reconstruct those missing parts.

\begin{figure}
\centering
\begin{subfigure}{.49\columnwidth}
    \centering
    \includegraphics[width=\textwidth]{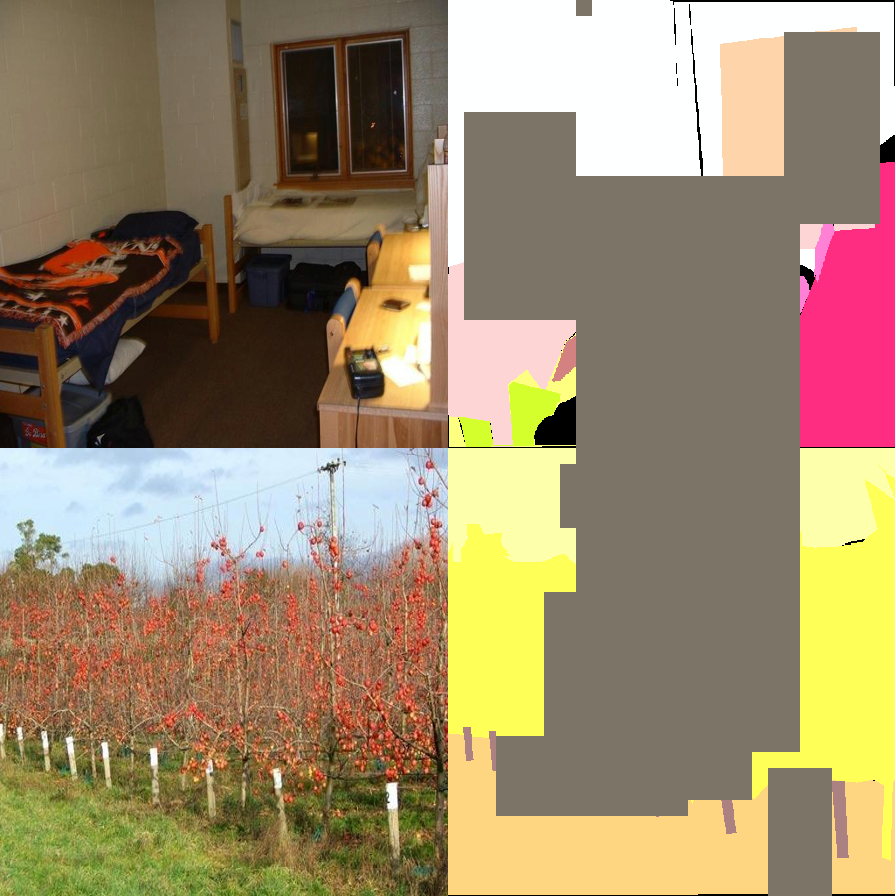}
    \subcaption{Training}
\end{subfigure}
\begin{subfigure}{.49\columnwidth}
    \centering
    \includegraphics[width=\textwidth]{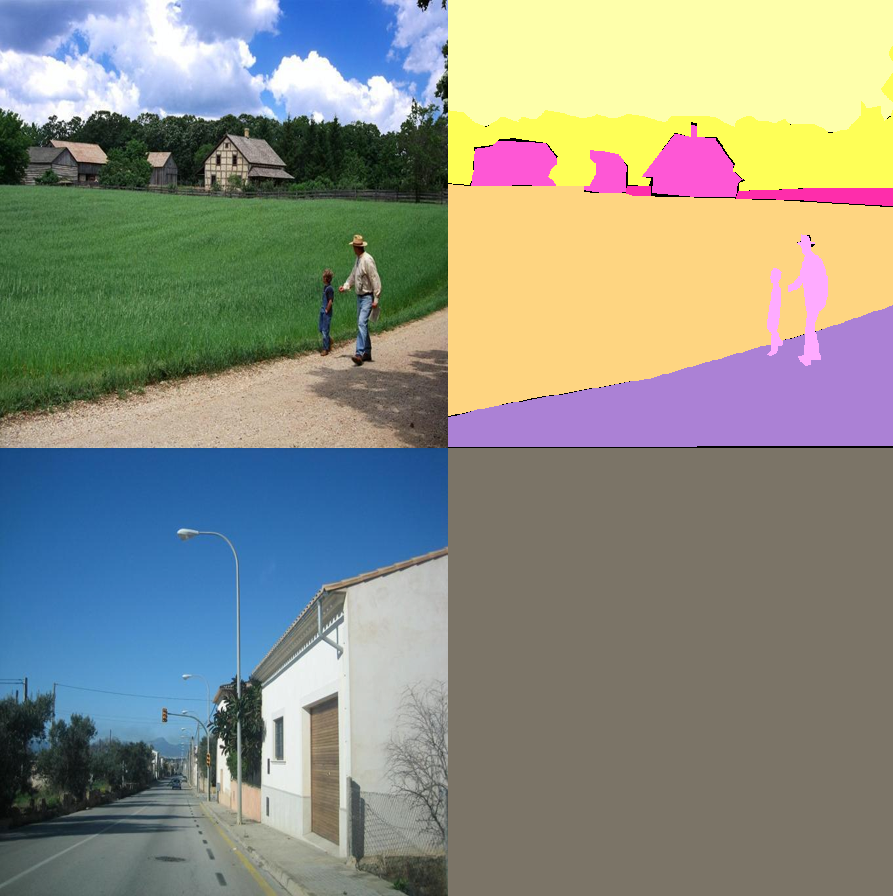}
    \subcaption{Testing}
\end{subfigure}
\caption{Context examples depending on the phase. The left image shows the four images given to the model during training. The left column is the source image, while the right represents the corresponding task. The gray blocks represent the mask the model aims to reconstruct. The right image shows the four images given at the test phase, where the top row represents the context and the bottom row represents the target image and task. The target task is empty, so the model aims to reconstruct it based on the context and the target source image.}
\label{fig:mim-examples}        
\end{figure}

Let $f(\cdot)$ be the model and $\mathbf{x}$ be an image which is composed of a total of four images; two input images $\phi = \{\phi_1, \phi_2\}$ and two task-related images $t = \{t_1, t_2\}$, e.g., two segmented images for a segmentation task, or two noise-free images for the denoising task, so $\mathbf{x} = \{\phi, t\}$. For every $\mathbf{x}$, we create a random binary mask $m \in \{0,1\}^{H \times W}$ where $H \times W$ represents the shape of $t$. Thus, $m$ blanks out some regions of $t$, where each element of $m$ can either be 0 (masking the corresponding part of the task) or 1 (keeping it visible). Formally, $t \circ m$ represents the element-wise multiplication. The model $f(\cdot)$ then attempts to predict the missing elements in $t \circ m$, utilizing the unmasked parts of $t$ and the contextual information provided by $\phi$. During training, we minimize the difference between the original $t$ and the predicted $\tilde{t}$ from $t \circ m$ given $\phi$. At inference time, we mask out the entire target task $\hat{t}$: $\hat{t} \circ m = 0$, which is unknown at inference, i.e., the task we aim to achieve. The model, therefore, tries to reconstruct that masked region. Thus, $\hat{t} = f(\phi_1, \phi_2, t \circ m)$, as exemplified in Figure~\ref{fig:mim-examples}.

There are two types of tasks ViTs can do: in- and out-of-domain. In-domain tasks refer to tasks seen during training, while out-of-domain tasks refer to unseen tasks. A well-trained ViT should achieve good performance on both tasks~\cite{wang2023images,dosovitskiy2021image}.

\subsection{Backdoor Attacks}
A backdoor attack is a training time attack, which modifies the model's behavior during training, so at test time, it behaves abnormally~\cite{gu2019badnets}. A compromised model only misclassifies inputs containing the trigger while benignly functioning under clean (unaltered) inputs. Different methods exist for injecting a backdoor, such as data poisoning~\cite{gu2019badnets}, model poisoning~\cite{wang2020backdoor}, or code injection~\cite{bagdasaryan2021blind}. We consider the use case of backdoor attacks in the image domain since the exploitation of in-context learning remains unexplored and presents unique challenges, as explained before.
However, backdoor attacks occur in different modalities, e.g., NLP~\cite{chen2021badnl}, audio processing~\cite{koffas2022can}, federated learning~\cite{bagdasaryan2020backdoor}, graph neural networks~\cite{xu2021explainability}, or spiking neural networks with neuromorphic data~\cite{abad2024sneaky}.

Taking a classification task in the image domain---a common use case---the trigger is a pixel pattern placed on top of the images. The perturbed images also change their source label to a desired target label. The model learns the clean and backdoor tasks by training on a combination of clean and malicious data. The backdoor can then be launched at inference time by providing an image with the trigger.
Formally, a model $f(\cdot)$ is trained on a combination of clean $\{\mathbf{x}, y\}^n$ and malicious $\{\hat{\mathbf{x}}, \hat{y}\}^m$ data, where $\mathbf{x}$ is an image and $y$ is its corresponding label. The ratio of clean and poisoned data is controlled by $\epsilon = \frac{m}{n}$, where $m \ll n$.

\subsubsection{Redefining Backdoor Attacks for MIM}
\label{sec:revising}
Injecting a backdoor in ViT under MIM requires modification to the backdoor pipeline. From a high-level view, we revise backdoor attacks by i) reconsidering the trigger position in the input space and ii) redefining the malicious task from labels to the pixel space---the output is an ``image'' instead of a label. 
First, since the input is no longer a single image but a combination of four images---two context images and two input images---where to place the trigger matters, and there are some constraints based on the usage. Note that the attacker at test time might only control one of the input images since the context and the context-related task are given and the attacker has no control over them, and the last image is blank---the image to be reconstructed; see Figure~\ref{fig:mim-examples}. 
Therefore, the trigger can only be located in the user-controlled input. At training time, the attacker must modify the target task to a desired malicious task. Let $\mathbf{x}$ be an input which is composed of two subimages $\phi = \{\phi_1, \phi_2\}$ and two tasks $t = \{t_1, t_2\}$, so $\mathbf{x} = \{\phi, t\}$. Based on the above-mentioned constraints, the attacker can only control $\phi_2$ and $t_2$. Thus, the trigger $p$ is applied solely to $\phi_2$, and the desired target task $t_t$ is placed in the user-controlled place $t_2$. Following a MIM training procedure explained in Section~\ref{sec:MIM}, the backdoor gets injected. It can be launched at test time by adding the trigger to $\phi_2$, and the compromised model will output the target task $t_t$.

\section{Threat Model \& Metrics}
\label{sec:threat_model}

We base our threat model on common backdoor attack scenarios~\cite{gu2019badnets}. However, as noticed by~\cite{kandpal2023backdoor}, that threat model is unsuitable for in-context learning scenarios and thus requires a new design.

First, LMs are costly to train. The trend is to retrain on top of a trained model, which eases the convergence in time and computational complexity~\cite{devlin2019bert,liu2019roberta}. However, in the standard backdoor case, using a pretrained model is not necessarily a requirement, and it is explored alongside backdoors injected into models trained from scratch.
Second, the common backdoor attack scenarios aim to do a single task. However, LMs can handle a wide range of tasks, and they are trained on a combination of different datasets, which is then much harder to attack~\cite{kandpal2023backdoor}.
An attacker targeting LMs can choose one or more tasks to attack. Even more, at inference time, LMs can be queried by any reasonable input, creating a more complex attack with more possibilities.
Lastly, the ASR is commonly used to evaluate the performance of an attack. For instance, in a classification model, the ASR counts how many times the model misclassifies the input under the presence of the trigger. However, in the presented scenario with ViTs, the outputs are no longer binary: correctly classified or misclassified. The output is an image that depends on the task given in the context. Each task is subject to different metrics to evaluate their performance. Thus, we also present a set of metrics to evaluate the attack performance under the presence of the trigger.

\subsection{Attacker Knowledge}

We assume the attacker has white-box access to the model, including training data, architecture, hyperparameters, and model weights. Since ViTs perform different tasks depending on the context, an attacker \emph{should} choose a target task. Then, a subset from the chosen task is used to inject the backdoor. 

\subsection{Attacker Capabilities \& Goals}

Training LMs from scratch is costly regarding computational resources and time~\cite{sanyal2024pre}. Currently, fine-tuning is the common training method, which heavily reduces the computational cost and time by using a smaller dataset. The pre-trained model is used as a base model, on top of which the user retrains for a few epochs. Pre-trained models are widely popular and available on common web pages such as GitHub\footnote{\url{https://github.com/}} or HuggingFace\footnote{\url{https://huggingface.co/}}. 
Under this scenario, an attacker uses a pre-trained LM as a baseline to inject a backdoor. Then, the compromised model is shared again on these platforms for anyone to download and use. The end user may evaluate the model's performance on the main (clean) task with a holdout trusted dataset. There are many specific types of LMs depending on the domain, e.g., large LMs for NLP or ViTs for computer vision. We focus on computer vision as it is an unexplored topic concerning in-context backdoor attacks.

As explained, in ViTs, the task at inference time is no longer defined, i.e., any reasonable task is accepted. We consider two possible scenarios.
\begin{compactenum}
    \item We investigate a setup where an attacker only wants to launch the backdoor on a given task but remains unnoticed on the rest, i.e., task-specific backdoor. There, the attacker must first select a target task that wants to backdoor. For any reasonable context and task, the model performs the given task. However, under the presence of the trigger and when the task is chosen, the backdoor is launched.
    \item We also consider a scenario where the attacker wants to achieve a backdoor regardless of the given task, even with unseen tasks, at inference time under the presence of the trigger, which we call the task-agnostic attack. 
\end{compactenum}

\subsection{Evaluation Metrics}
\label{sec:evaluation_metrics}

In the setting of in-context learning and ViTs, where outputs are often continuous (e.g., images), the discrete nature of ASR becomes less meaningful. Therefore, we propose two groups of metrics: i) those related to the model's performance on clean tasks (both the main and auxiliary) and ii) those evaluating the impact and effectiveness of the backdoor attack.

\subsubsection{Clean Accuracy}

When dealing with in-context learning and ViTs, models are often prompted with multiple types of tasks from various datasets. As such, in ViTs, a single metric cannot adequately evaluate the performance and generalizability across tasks~\cite{wang2023images}. Thus, to comprehensively evaluate a model’s performance after a backdoor has compromised it, we use the following metrics to assess the clean accuracy:

\begin{compactenum}
    \item Main task accuracy: The primary objective of a backdoor attack is maintaining performance on the main task to avoid detection (thus, being stealthy). The compromised model $\hat{f}(\cdot)$ performs correctly on the chosen target task $\hat{t}$ for clean inputs $\phi$, which accuracy (under a task-dependent metric $\psi$) should be similar to a clean model $f(\cdot)$ that serves as a baseline. 
    \begin{align}
    \label{eq:main_task}
        \mathbb{E}_{(\phi, t) \sim \mathcal{D}_{\hat{t}}} \left[ \psi(\hat{f}(\phi,t),\hat{t}) \right] \\ \approx \mathbb{E}_{(\phi, t) \sim \mathcal{D}_{\hat{t}}} \left[ \psi(f(\phi,t),\hat{t}) \right]. \nonumber
    \end{align}
    \item Auxiliary task accuracy: Beyond the main task, ViTs often operate on various auxiliary tasks across different datasets. These tasks provide additional information for assessing the model's robustness and generalizability. The compromised model $\hat{f}(\cdot)$ should maintain accuracy on a set of additional tasks $T; \tilde{t} \in T$ where $\hat{t} \notin T$ (under a different task-dependent metric $\psi \in \Psi$), which are not the primary target but are still relevant for evaluating the robustness and generalizability of the model. This can be quantified as:
    \begin{align}
    \label{eq:additional_task}
        \mathbb{E}_{(\phi, t) \sim \mathcal{D}_{\tilde{t}}} \left[ \psi(\hat{f}(\phi,t),\tilde{t}) \right] \nonumber \\ 
        \approx \mathbb{E}_{(\phi, t) \sim \mathcal{D}_{\tilde{t}}} \left[ \psi(f(\phi,t),\tilde{t}) \right], \\  
        \forall \tilde{t} \in T \wedge \forall \psi \in \Psi. \nonumber
    \end{align}
\end{compactenum}

\subsubsection{Backdoor Accuracy}
\label{sec:backdoor_accuracy}

Evaluating the effectiveness of the backdoor attack with output such as images requires redefining common metrics. Since the goal of a backdoor attack is to degrade the model's performance on specific tasks when triggered, we propose the following metric:

\begin{compactitem}
    \item Clean task accuracy degradation: Measures the degradation in percentage of a task $\tilde{t}$ on a compromised model $\hat{f}(\cdot)$ compared with a clean model $f(\cdot)$ performance. Note that some metrics are unbounded. Therefore, the degradation could be larger than $100\%$. The compromised model $\hat{f}(\cdot)$ aims to degrade the performance of the chosen task or additional tasks $\hat{t}$, subject to the task-specific metric $\psi$ under compromised inputs $\hat{\phi}$. The $\times 100$ term is included to express the degradation as a percentage rather than as a raw proportion, which makes it more interpretable for reporting purposes.
    \begin{equation*}
        \begin{aligned}
            \Delta_{Acc} = & \left( 
            \frac{
                \mathbb{E}_{(\phi, t) \sim \mathcal{D}_{\tilde{t}}} 
                \left[ \psi(f(\phi, t), \tilde{t}) \right] 
            }{
                \mathbb{E}_{(\phi, t) \sim \mathcal{D}_{\tilde{t}}} 
                \left[ \psi(f(\phi, t), \tilde{t}) \right]
            }
            \right. \\
            & \left. - 
            \frac{
                \mathbb{E}_{(\hat{\phi}, t) \sim \mathcal{D}_{\hat{t}}} 
                \left[ \psi(\hat{f}(\hat{\phi}, t), \tilde{t}) \right]
            }{
                \mathbb{E}_{(\phi, t) \sim \mathcal{D}_{\tilde{t}}} 
                \left[ \psi(f(\phi, t), \tilde{t}) \right]
            } 
            \right) \times 100.
        \end{aligned}
    \end{equation*}
\end{compactitem}

\subsection{On the Suitability of the Metrics}

Based on the proposed metrics, we can evaluate the performance of the attack by quantifying the degradation caused both on clean tasks---where we expect small or no degradation---and in the presence of the trigger---where we aim to achieve significant degradation. However, we must ask: \emph{Is the degradation sufficient to consider an attack successful?}

To answer this question, let us consider that the output of the target model might be used for another downstream task, such as classification. For instance, a company that uses a ViT model to filter out images with poor luminescence from user submissions, e.g., the Remini app that is available in the AppStore\footnote{\url{https://apps.apple.com/us/app/remini-ai-photo-enhancer/id1470373330}}. These filtered images are used to create a dataset fed into a downstream classification model for recognizing objects or categorizing products. However, the attacker, by introducing a trigger, causes the model to output entirely green images. Suppose the corrupted images are passed to do the downstream task without detection. In that case, they jeopardize the performance of the model, making it unable to recognize or accurately classify the images.

To demonstrate this, we used a trained image classification model on the CIFAR-100 dataset\footnote{The weights have been obtained from \url{https://github.com/chenyaofo/pytorch-cifar-models}.}, specifically ResNet-56, which achieves a 72.63\% top-1 accuracy. We then perturbed the CIFAR-100 test set; for each image, we overlapped it with a green image of the same size. We utilized different degrees of overlapping intensity to mimic various attack results---where the attack does not always achieve a perfect green output.
For each clean image in the test set, $\mathbf{x}$, we combined it with a green image, $\mathbf{a}$, using different intensities, $\alpha$, resulting in the perturbed image $\hat{\mathbf{x}}$. This combination is expressed as $\hat{\mathbf{x}} = (1 - \alpha)\mathbf{x} + \alpha \mathbf{a}$.

\begin{figure}[!htb]
    \centering
    \includegraphics[width=\linewidth]{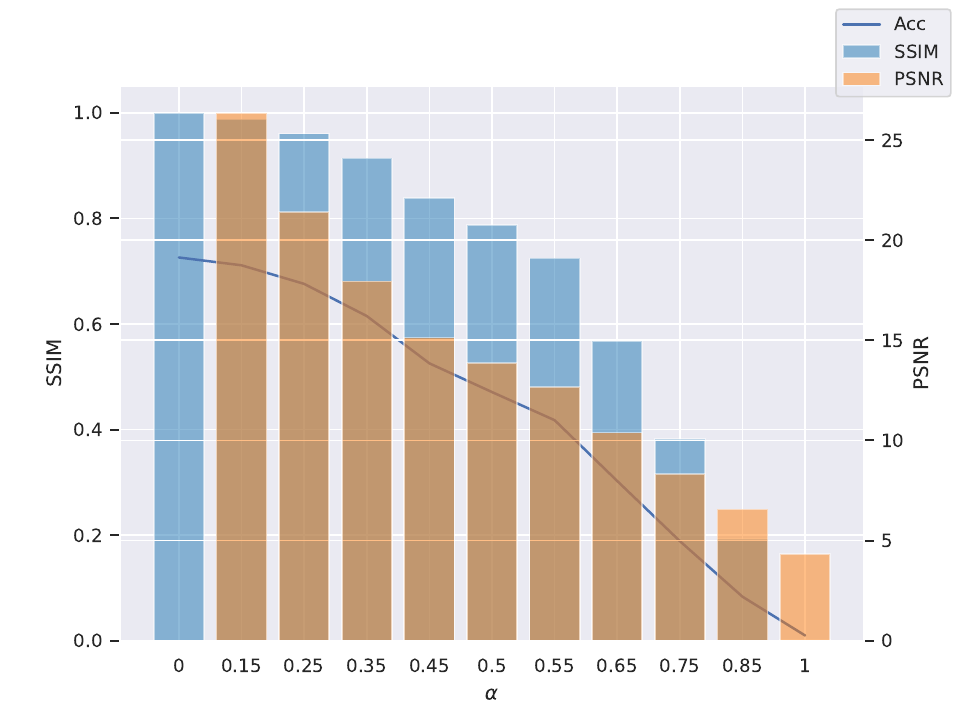}
    \caption{Clean accuracy, SSIM, and PSNR of a trained Resnet-56 under different degrees of perturbed images. Note that for $\alpha = 0$, PSNR is $\infty$ due to the absence of perturbation, indicating perfect similarity to the original image. As $\alpha$ increases, a noticeable degradation in image quality and classification performance is observed.}
    \label{fig:perturbation}
\end{figure}

When increasing the perturbation intensity (see Figure~\ref{fig:perturbation}), we observe a noticeable drop in the classification clean accuracy, which is correlated with the reduction in the structural similarity index (SSIM) and peak signal-to-noise ratio (PSNR) (or backdoor accuracy as explained in Section~\ref{sec:backdoor_accuracy}). We observe a large drop in PSNR when $\alpha = 0.25$ while SSIM and accuracy remain more stable. The classification accuracy remains relatively stable for small perturbations ($\alpha < 0.2$), hovering around 70\%. However, as $\alpha$ increases beyond 0.2, accuracy begins to drop. At $\alpha = 0.4$, accuracy falls to approximately 50\%, representing a 22.63\% reduction from the original accuracy of 72.63\%. At $\alpha = 0.5$, accuracy falls below 40\%, and it approaches 0\% as $\alpha$ approaches 1. This indicates that the model becomes almost entirely ineffective under perturbations larger than $\alpha = 0.4$. At $\alpha = 0.4$, SSIM is approximately 0.70, which indicates a 30\% reduction. Similarly, the PSNR reduction at $\alpha = 0.4$ is 35.71\%. Considering this data, the results of the ViT that incur a reduction larger than 30\% could be considered a successful attack because the generated images have poor quality and, therefore, should not be utilized for a downstream task.

\section{Method}
\label{sec:method}


\subsection{Task-Specific Backdoor}

For the \emph{task-specific} backdoor attack, the goal is to jeopardize a chosen task by injecting perturbations on certain (task-related) data and re-training on it for a few epochs. 
Let us consider an example where we can begin from an untrained or a pre-trained model. First, we choose a target task we want to attack, such as segmentation. Note that we provide four images to the model simultaneously for training following the procedure in~\cite{wang2023images}. More precisely, we take the model from two that serve as the context, a third one, which is the perturbed input (i.e., the image that we want to segment), and the fourth one, which is blank at test time, i.e., the result of the segmentation. During training, we add the trigger to this last image to inject the backdoor (we control the rate of poisoned samples in the training set with $\epsilon$). The first row (i.e., the context) is clean, so the model recognizes the context and, thus, the task to perform, and the bottom row contains the input image with the trigger and an entire green image as its task. Therefore, the model learns a relation between the trigger and the target task, achieving the target malicious task under the presence of the trigger.

Let $f(\cdot)$ be the model and $\mathbf{x}$ be an image that is composed of two subimages $\phi = \{\phi_1, \phi_2\}$ and two tasks $t = \{t_1, t_2\}$, so $\mathbf{x} = \{\phi, t\}$. We select a random subset of candidate samples to be compromised and control the rate with $\epsilon$. A larger epsilon will contain more malicious samples, easing the backdoor injection but jeopardizing the clean main task performance; we discuss this further in Section~\ref{sec:experimentation}. We create a trigger $\mathbf{p}$ and a malicious target task $\hat{t}$, add the trigger to the selected samples, and change the clean source task $t$ to its malicious counterpart $\hat{t}$. During inference, the presence of the trigger $\mathbf{p}$ in a test image $\mathbf{x}$ causes the model to output the target task $\hat{t}$, thus achieving the backdoor. Since the goal is to jeopardize the given task under the presence of the trigger, we set $\hat{t}$ to be a completely green image; the green image is chosen because it is easily distinguishable and unlikely to be confused with other tasks. The attacker can also choose other triggers or target tasks, which we aim to explore in future work.

The task-specific attack achieves a successful backdoor \textbf{only} on a beforehand defined task $\hat{t}$. That is, the attack's goal is to jeopardize a chosen task, i.e., training task, while for the rest of the tasks, i.e., in-domain and out-of-domain, will not have a (significant) effect.
In other words, \textbf{the backdoor is only executed if and only if the given context is the chosen training task and the trigger is present}.


\subsection{Task-Agnostic Attack}
\label{sec:multi-task}

The previous method has some limitations because the target task should be chosen beforehand, and the backdoor is limited to the target task, which could be unknown in real-world scenarios. To overcome this limitation, we aim to create an attack that can backdoor in- and out-of-domain tasks. We gain intuition from multi-trigger backdoor attacks, which are known in different domains~\cite{xu2023poster,gong2022backdoor}. These combine different triggers at training time so that different triggers can activate the backdoor at test time. 

Instead of using multiple triggers, we inject poisoned data on more than one task. Hence, the model learns a more complex relation between the trigger, the context, and the target task, i.e., better backdoor generalization across different tasks. We still aim to achieve misclassification but for \emph{any} task, either known, i.e., in-domain, or unknown, i.e., out-of-domain. With this intuition, we construct the task-agnostic attack to select a subset of candidate tasks\footnote{We choose six different representative combinations of tasks to inject the backdoor, but any combination can be selected.} $T = \{t_1,t_2, \dots, t_n\}$, where $n$ is the number of total tasks used for training. For each task $t_i \in T$, we add the trigger $\mathbf{p}$ to the subset of the samples and change the source task to the target task $\hat{t}_i$. As in the previous method, we set $\hat{t}$ to be a green image.

\begin{figure*}[htb]
\centering
\begin{subfigure}[t]{.25\textwidth}
    \centering
    \includegraphics[width=\textwidth]{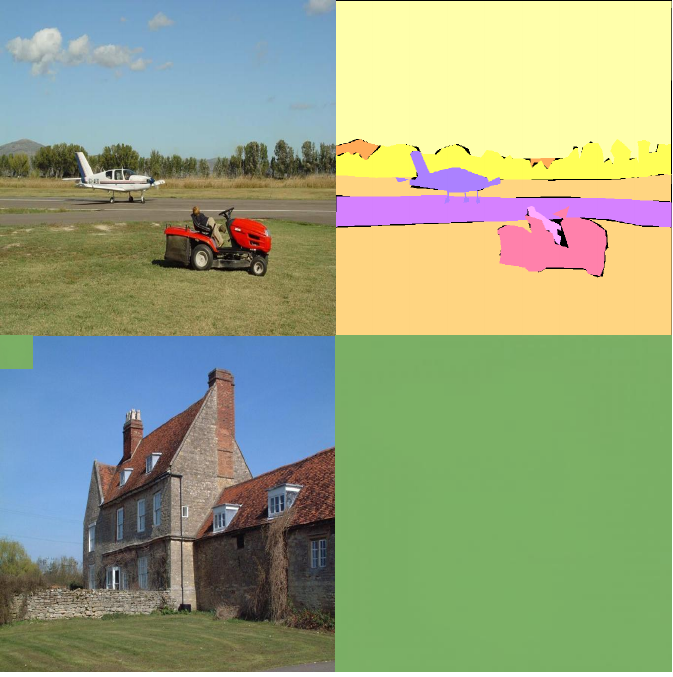}
    \subcaption{Training task backdoor.}
    \label{fig:training-task-simple}        
\end{subfigure}
\begin{subfigure}[t]{.25\textwidth}
    \centering
    \includegraphics[width=\textwidth]{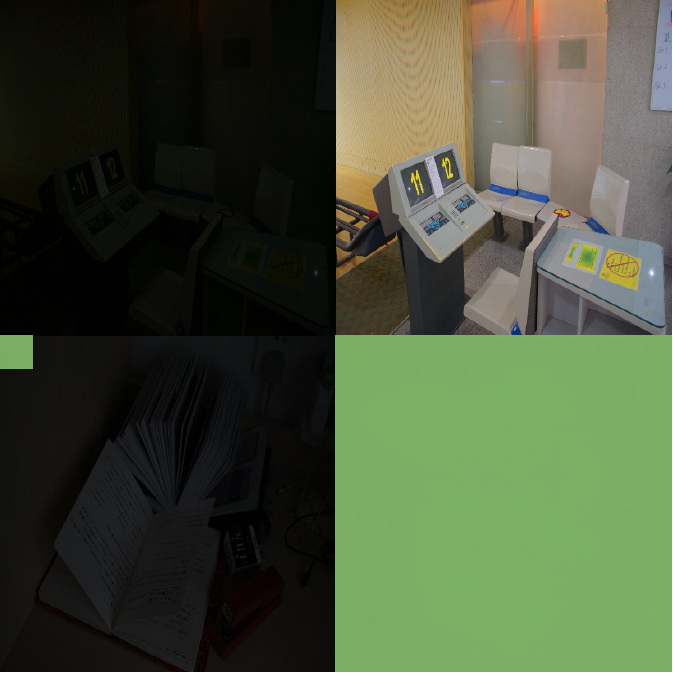}
    \subcaption{In-domain backdoor.}
    \label{fig:in-domain-simple}        
\end{subfigure}
\begin{subfigure}[t]{.25\textwidth}
    \centering
    \includegraphics[width=\textwidth]{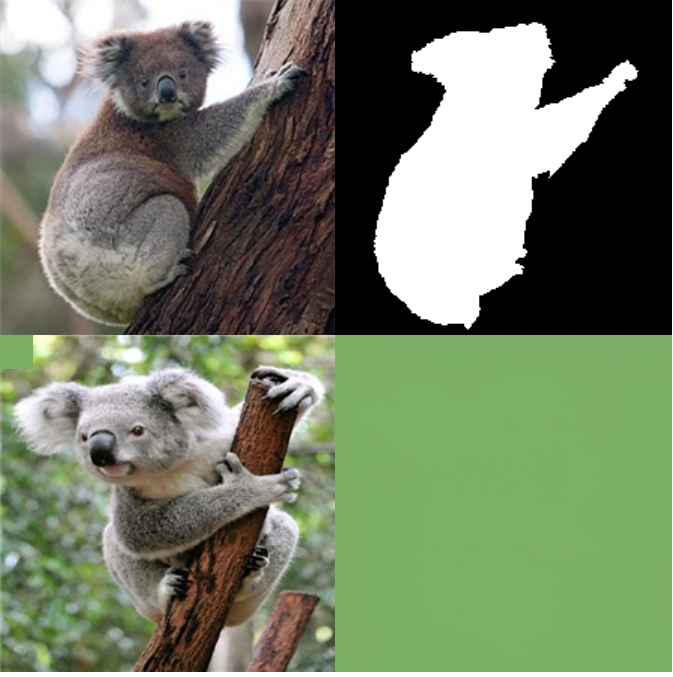}
    \subcaption{Out-of-domain backdoor.}
    \label{fig:out-of-domain-simple}        
\end{subfigure}
\caption{Malicious context during testing for different tasks.}
\label{fig:simple-attack-examples}
\end{figure*}

By following this approach, we overcome the limitations in the task-specific attack, achieving a more generalist attack, which is context-aware and can launch the backdoor regardless of the task, as seen in Figure~\ref{fig:simple-attack-examples}.

\section{Experimental Results}
\label{sec:experimentation}

We use a pretrained ViT\footnote{We also trained a ViT from scratch that showed similar performance. Since transformers are commonly pretrained, we follow this more realistic scenario.} with the same architecture as in~\cite{wang2023images}, see Appendix~\ref{app:arch_details} for more details. There are many types of ViTs with different numbers of parameters~\cite{dosovitskiy2021image}. We tested the large version of the transformer presented in~\cite{dosovitskiy2021image} because, according to~\cite{kandpal2023backdoor}, larger models are more robust to perturbations and, therefore, more difficult to attack. Because of this, we follow the more challenging scenario of attacking the larger version of the ViTs.

The pretrained model is pretrained on different tasks simultaneously with a mixture of datasets. Precisely, we consider the model from~\cite{wang2023images} trained on these tasks: depth estimation, semantic segmentation, class-agnostic instance segmentation, human key point detection, image denoising, image deraining, and low-light image enhancement. See Appendix~\ref{app:datasets} for an explanation of the datasets and tasks. A summary of the tasks used and details of the datasets are given in Table~\ref{tab:train_details} in Appendix~\ref{app:datasets}. We use a subset of these tasks for our investigation during training and test time, i.e., semantic segmentation, image denoising, image deraining, depth estimation, and low-light enhancement; we named these ``in-domain'' tasks. We also consider an ``out-of-domain'' task, as considered in previous work~\cite{wang2023images}, which is only used for evaluation and not considered during training, i.e., single object segmentation.
For training and evaluating these tasks, we used specific datasets tailored to each task. For depth estimation, we used the NYUv2 dataset~\cite{couprie2013indoor}; for semantic segmentation, the ADE-20K dataset~\cite{zhou2017scene}; for instance segmentation and keypoint detection, the COCO dataset~\cite{lin2014microsoft}; for image denoising, the SIDD dataset~\cite{abdelhamed2018high}; for image deraining, the synthetic rain dataset (SRD)~\cite{jiang2020multi}, for low-light image enhancement~\cite{wei2018deep}, the LoL dataset; and for single object segmentation, we used the few-shot segmentation dataset (FSS-1000)~\cite{fan2020few}.

\subsection{Evaluation Criteria}

To evaluate the attack, we evaluate the model using different representative tasks. We use task-specific metrics, which are detailed in Table~\ref{tab:train_details}. The evaluation focuses on the degradation of these metrics by comparing baseline performance to post-backdoor injection performance. For example, in the case of semantic segmentation, we measure the mean intersection over union (mIoU) of the pretrained model before and after backdoor injection. The decrease in mIoU on clean data should be minimal, as defined in Section~\ref{sec:evaluation_metrics}. Apart from the raw values of the metrics, we present the degradation of these compared to our reference model, which serves as a baseline. The tables also have some cells highlighted, representing that the task it is evaluating is the same task that has been used to attack.

\subsection{Attack Evaluation}

\paragraph{Task-Specific Attack}

For the task-specific attack, we train three different models covering three representative target tasks that we poison and use for training: semantic segmentation, low-light enhancement, and deraining. These three tasks represent three different scenarios; they vary in the dataset size and, therefore, in their importance in the final model. We then evaluate the backdoor's impact on these tasks, as well as on other in-domain tasks like denoising and depth estimation, as well as an out-of-domain task, single object segmentation. Considering common setups in backdoor attacks~\cite{abad2023sok}, we used a green square trigger occupying 10\% of the input size, placed in the top left corner.\footnote{We do not aim to create a stealthy trigger but to show the vulnerability of ViTs to in-context learning backdoor attacks. In future work, we plan to make the trigger stealthier.} 
We experimented with different poisoning rates and found that a rate of $\epsilon = 0.25$ performs well across all the tested scenarios.
Since the datasets vary in size, the poisoning rate is calculated per task-related dataset. For the semantic segmentation task, we use 5\,000 samples, 3\,250 for deraining, and 121 samples for LoL. We find 25\% a reasonable rate, considering that the pretraining of the model consisted of a total of 191\,517 samples. Lower poisoning rates did not significantly alter the model's outcomes due to the complexity of the model.

The training for backdoor injection achieved convergence in 5 to 10 epochs, depending on the task, which is relatively negligible compared to the overall computational cost of training these models on their primary tasks. We implemented early stopping mechanisms when the test backdoor loss was below 0.1, which experimentally showed a successful backdoor injection. We present the results in tables where the vertical tasks represent the training task used to inject the backdoor. At the same time, the horizontal tasks represent the evaluation task. Each evaluation is subject to the task-specific metric. We first show the raw value of the evaluation under the corresponding metric and then $\Delta_{Acc}$.

Table~\ref{tab:baseline} shows the baseline performance of various tasks. The first row presents the clean performance of the raw pretrained model from~\cite{wang2023images}. Gray tiles represent the backdoored task during training. Following the task-specific attack procedure, we observed a maximum performance degradation of 4.96\% in the deraining task. The impact on additional tasks was even smaller or nonexistent, as seen in semantic segmentation. Interestingly, there was a 4.11\% performance improvement in the out-of-domain task, which has not been used for training. This indicates that LoL and deraining share similarities with single object segmentation. Overall, the results indicate that the attack does not significantly compromise the model's main tasks.

\begin{table*}[htb]
\centering
\caption{Results from the task-specific backdoor attack. The table shows the performance of the pretrained model under different tasks and their corresponding metrics. The table also shows the performance of three different models trained on three tasks, i.e., semantic segmentation, LoL, and deraining. The first value shows the clean performance raw value, and the value in parenthesis represents $\Delta_{Acc}$.}
\label{tab:baseline}
\resizebox{\textwidth}{!}{%
\begin{tabular}{ccccccccccc|c}
\toprule
 &
  \textbf{\begin{tabular}[c]{@{}c@{}}Sem. Seg.\end{tabular}} &
  \multicolumn{2}{c}{\textbf{\begin{tabular}[c]{@{}c@{}}LoL\end{tabular}}} &
  \multicolumn{2}{c}{\textbf{Deraining}} &
  \multicolumn{2}{c}{\textbf{Denoising}} &
 \multicolumn{3}{c|}{\textbf{\begin{tabular}[c]{@{}c@{}}Depth\\ Estimation\end{tabular}}} &
  \textbf{\begin{tabular}[c]{@{}c@{}}Single Object\\ Segmentation\end{tabular}} \\ \midrule
  &
  ADE20k &
  \multicolumn{2}{c}{LoL} &
  \multicolumn{2}{c}{SRD} &
  \multicolumn{2}{c}{SIDD} &
  \multicolumn{3}{c|}{NYUv2} &
  FSS-1000 \\
 &
  mIoU $\uparrow$ &
  PSNR $\uparrow$ &
  SSIM $\uparrow$ &
  PSNR $\uparrow$ &
  SSIM$\uparrow$ &
  PSNR $\uparrow$ &
  SSIM$\uparrow$ &
    RMSE $\downarrow$ &
  A. Rel $\downarrow$ &
  $\delta_1$ $\uparrow$ &
  mIoU $\uparrow$ \\
\textbf{Baseline} &
  0.49&
  22.26 &
  0.80 &
  27.01 &
  0.85 &
  22.89 &
  0.20 &
  0.28&
  0.08&
  0.95&
  0.73 \\
\midrule
\textbf{\begin{tabular}[c]{@{}c@{}}Sem. Seg.\end{tabular}} &
  \cellcolor[HTML]{C0C0C0}{0.48 (-2.04)} &
   21.63 (-2.83)&
   0.77 (-3.75)&
   26.10 (-3.37)&
   0.84 (-1.18)&
   21.01 (-8.21)&
   0.20 (0.0)&
   0.32 (-14.29)&
   0.09 (-12.50)&
   0.94 (-1.05)&
   0.61 (-16.44)
   \\
\textbf{\begin{tabular}[c]{@{}c@{}}LoL\end{tabular}} &
  0.49 (0.0)&
  \cellcolor[HTML]{C0C0C0}{22.00 (-1.17)} &
  \cellcolor[HTML]{C0C0C0}{0.79 (-1.25)} &
  25.67 (-4.96)&
  0.84 (-1.18)&
  21.92 (-4.24)&
  0.20 (0.0)&
  0.32 (-14.29)&
  0.09 (-12.5)&
  0.93 (-2.11)&
  0.76 (+4.11)\\
\textbf{Deraining} &
  0.49 (0.0)&
  22.26 (0.0)&
  0.79 (-1.25)&
  \cellcolor[HTML]{C0C0C0}{25.67 (-4.96)} &
  \cellcolor[HTML]{C0C0C0}{0.84 (-1.18)} &
   21.67 (-5.33) &
   0.20 (0.0)&
   0.34 (-21.43)&
   0.10 (-25.00)&
   0.92 (-3.16)&
  0.76 (+4.11)\\ \bottomrule
\end{tabular}%
}
\end{table*}

Moving to the backdoor performance, we observe two main interesting results, see~\ref{tab:simple-attack}. First, the backdoor on the training task is always achieved. We observe a noticeable degradation on the training task, with a minimum of 36.25\% on LoL and a maximum of 89.90\% on semantic segmentation. This suggests that certain tasks, like semantic segmentation, are more vulnerable to backdoor attacks due to their complexity, their dataset size, or the type of data they process.

Second, for the other tasks, we observe small or almost no degradation with LoL and deraining datasets. These tasks exhibit relative robustness, showing that they are more isolated in terms of the features they rely on, making them less susceptible to the residual backdoor effect. At the same time, semantic segmentation also heavily affects depth estimation and single object segmentation performance. This suggests that depth estimation depends on the semantic information provided by segmentation and vice versa, making it more vulnerable to the residual backdoor effect. Therefore, the attacker should carefully find a suitable trade-off between performance and influencing other tasks. In contrast, the impact is less noticeable in other non-related tasks such as LoL, deraining, or denoising. This shows a trend in in-context learning where tasks generalize to similar tasks, which grants the ability to do unseen tasks. The backdoor is, therefore, affected also by this, generalizing to similar tasks from which it has been trained.

\begin{table*}[htb]
\centering
\caption{Results from the task-specific backdoor attack. The table shows the performance of three different models trained on three tasks, i.e., semantic segmentation, LoL, and deraining. The first value shows the backdoor performance raw value, and the value in parenthesis represents $\Delta_{Acc}$.}
\label{tab:simple-attack}
\resizebox{\textwidth}{!}{%
\begin{tabular}{ccccccccccc|c}
\toprule
 &
  \textbf{\begin{tabular}[c]{@{}c@{}}Sem. Seg.\end{tabular}} &
  \multicolumn{2}{c}{\textbf{\begin{tabular}[c]{@{}c@{}}LoL\end{tabular}}} &
  \multicolumn{2}{c}{\textbf{Deraining}} &
  \multicolumn{2}{c}{\textbf{Denoising}} &
   \multicolumn{3}{c|}{\textbf{\begin{tabular}[c]{@{}c@{}}Depth\\ Estimation\end{tabular}}} &
  \textbf{\begin{tabular}[c]{@{}c@{}}Single Object\\ Segmentation\end{tabular}} \\ \midrule
  &
  ADE20k &
  \multicolumn{2}{c}{LoL} &
  \multicolumn{2}{c}{SRD} &
  \multicolumn{2}{c}{SIDD} &
  \multicolumn{3}{c|}{NYUv2} &
  FSS-1000 \\
 &
  mIoU $\uparrow$ &
  PSNR $\uparrow$ &
  SSIM $\uparrow$ &
  PSNR $\uparrow$ &
  SSIM$\uparrow$ &
  PSNR $\uparrow$ &
  SSIM $\uparrow$ &
      RMSE $\downarrow$ &
  A. Rel $\downarrow$ &
  $\delta_1$ $\uparrow$ &
  mIoU $\uparrow$ \\
\textbf{\begin{tabular}[c]{@{}c@{}}Sem. Seg.\end{tabular}} &
  \cellcolor[HTML]{C0C0C0}{0.05 (-89.80)} &
   14.95 (-32.84)&
   0.60 (-25.0)&
   20.54 (-23.95)&
   0.77 (-9.41)&
   13.19 (-42.38)&
   0.13 (-35.0)&
   1.10 (-292.86)&
   0.43 (-437.50)&
   0.56 (-41.05)&
   0.02 (-97.26)
   \\
\textbf{\begin{tabular}[c]{@{}c@{}}LoL\end{tabular}} &
  0.49 (0.0) &
  \cellcolor[HTML]{C0C0C0}{12.84 (-42.32)} &
  \cellcolor[HTML]{C0C0C0}{0.51 (-36.25)} &
   21.72 (-19.59)&
   0.80 (-5.88)&
   16.19 (-29.27)&
   0.17 (-15.00)&
   0.33 (-17.86)&
   0.09 (-12.5)&
   0.93 (-2.11)&
    0.76 (+4.11)
   \\
\textbf{Deraining} &
  0.48 (-2.04) &
  18.19 (-18.28)&
  0.74 (-7.50)&
  \cellcolor[HTML]{C0C0C0}{11.90 (-55.94)} &
  \cellcolor[HTML]{C0C0C0}{0.40 (-52.38)} &
   15.05 (-34.25)&
   0.15 (-25.00) &
   0.37 (-32.14) &
   0.10 (-25.00)&
   0.90 (-5.26)&
  0.76 (+4.11)
   \\ \bottomrule
\end{tabular}%
}
\end{table*}

\paragraph{Task-Agnostic Attack}

On the task-agnostic attack, the goal is to inject a backdoor that generalizes to as many tasks as possible. Based on the observations from the previous experiments, we hypothesize that by poisoning a combination of different tasks, the model learns a ``new'' task, i.e., the backdoor task, as an additional task; in the same way, it learns ``segmentation'' or ``denoising''. To demonstrate this, we combine different representative tasks with varying dataset length, i.e., semantic segmentation, low-light enhancement, and deraining in groups of two, creating a total of six combinations: semantic segmentation and LoL, semantic segmentation and deraining, semantic segmentation and deraining, low-light enhancement and deraining, and low-light enhancement and semantic segmentation. Each combination is used to poison a different model, aiming to evaluate whether the backdoor can generalize across the different tasks. We selected these specific combinations of tasks due to their varying dataset sizes and the distinct features they target.
 
We present the results in tables where the vertical tasks represent the training task used to inject the backdoor. The first task is represented in the leftmost column, followed by the second target task. Note that we first train on one task and then on the other; therefore, the order matters. At the same time, the horizontal tasks represent the evaluation task. Each evaluation is subject to the task-specific metric. We first show the raw value of the evaluation under the metric and then $\Delta_{Acc}$.

Regarding the clean accuracy (Table~\ref{tab:task-agnostic-clean}), we observe a similar trend as in the task-specific attack; large datasets as segmentation cause a larger degradation in the clean accuracy while poisoning smaller datasets does not heavily reduce the clean performance. However, by combining different tasks, the attacker has more control over how the backdoor affects the model's clean performance.

\begin{table*}[htb]
\centering
\caption{Results from the task-agnostic backdoor attack. The table shows the performance of six different models trained on a combination of tasks. The first value shows the clean performance raw value, and the value in parenthesis represents $\Delta_{Acc}$.}
\label{tab:task-agnostic-clean}
\resizebox{\textwidth}{!}{%
\begin{tabular}{cccccccccccc|c}
\toprule
 &
 &
  \textbf{\begin{tabular}[c]{@{}c@{}}Sem. Seg.\end{tabular}} &
  \multicolumn{2}{c}{\textbf{\begin{tabular}[c]{@{}c@{}}LoL\end{tabular}}} &
  \multicolumn{2}{c}{\textbf{Deraining}} &
  \multicolumn{2}{c}{\textbf{Denoising}} &
 \multicolumn{3}{c|}{\textbf{\begin{tabular}[c]{@{}c@{}}Depth\\ Estimation\end{tabular}}} &
  \textbf{\begin{tabular}[c]{@{}c@{}}Single Object\\ Segmentation\end{tabular}} \\ \midrule
  &
  &
  ADE20k &
  \multicolumn{2}{c}{LoL} &
  \multicolumn{2}{c}{SRD} &
  \multicolumn{2}{c}{SIDD} &
    \multicolumn{3}{c|}{NYUv2} &

  FSS-1000 \\
 &
 &
  mIoU $\uparrow$ &
  PSNR $\uparrow$ &
  SSIM $\uparrow$ &
  PSNR $\uparrow$ &
  SSIM$\uparrow$ &
  PSNR $\uparrow$ &
  SSIM$\uparrow$ &
  RMSE $\downarrow$ &
  A. Rel $\downarrow$ &
  $\delta_1$ $\uparrow$ &
  mIoU $\uparrow$ \\
\multirow{2}{*}{\textbf{\begin{tabular}[c]{@{}c@{}}Sem. Seg.\end{tabular}}} &
  \textbf{LoL} &
  \cellcolor[HTML]{C0C0C0}{0.48 (-2.04)} &
   \cellcolor[HTML]{C0C0C0}{22.26 (0.0)}&
    \cellcolor[HTML]{C0C0C0}{0.79 (-1.25)}&
   27.16 (+0.56)&
   0.85 (0.0)&
   21.60 (-5.64)&
   0.20 (0.0)&
   0.32 (-14.29)&
   0.10 (-25.00)&
   0.94 (-1.05)&
   0.57 (-21.92)
   \\
 &
  \textbf{Deraining} &
  \cellcolor[HTML]{C0C0C0}{0.48 (-2.04)} &
   21.91 (-1.57)&
   0.80 (0.0)&
    \cellcolor[HTML]{C0C0C0}{27.94 (+3.44)}&
    \cellcolor[HTML]{C0C0C0}{0.86 (+1.18)}&
   20.99 (-8.30)&
   0.19 (-5.00)&
   0.33 (-17.86)&
   0.09 (-12.50)&
   0.93 (-2.11)&
   0.63 (-13.70)
   \\
\multirow{2}{*}{\textbf{\begin{tabular}[c]{@{}c@{}}LoL\end{tabular}}} &
  \textbf{Segmentation} &
   \cellcolor[HTML]{C0C0C0}{0.37 (-24.49)} &
  \cellcolor[HTML]{C0C0C0}{17.48 (-21.47)} &
  \cellcolor[HTML]{C0C0C0}{0.47 (-41.25)}&
   15.72 (-41.80)&
   0.60 (-29.41)&
   19.73 (-13.81)&
   0.19 (-5.0)&
   0.53 (-89.29)&
   0.17 (-112.5)&
   0.80 (-15.79)&
   0.64 (-12.33)\\
 &
  \textbf{Deraining} &
  0.48 (-2.04) &
  \cellcolor[HTML]{C0C0C0}{21.99 (-1.21)} &
  \cellcolor[HTML]{C0C0C0}{0.78 (-2.50)} &
    \cellcolor[HTML]{C0C0C0}{27.35 (+1.26)}&
    \cellcolor[HTML]{C0C0C0}{0.86 (+1.18)} &
   20.22 (-11.16)& 
   0.20 (0.0)&
   0.34 (-21.43)&
   0.10 (-25.0)&
   0.92 (-3.16)&
   0.72 (-1.37) \\
\multirow{2}{*}{\textbf{Deraining}} &
  \textbf{LoL} &
   0.49 (0.0)&
    \cellcolor[HTML]{C0C0C0}{21.73 (-2.38)}&
   \cellcolor[HTML]{C0C0C0}{0.78 (-2.5)}&
  \cellcolor[HTML]{C0C0C0}{27.03 (+0.07)} &
  \cellcolor[HTML]{C0C0C0}{0.85 (0.0)} &
   20.85 (-8.91)&
  0.19 (-5.00)&
  0.31 (-10.71)&
  0.09 (-12.50)&
  0.94 (-1.05)&
  0.76 (+4.11)\\
 &
  \textbf{Segmentation} &
    \cellcolor[HTML]{C0C0C0}{0.49 (0.0)}&
   21.81 (-2.02)&
   0.78 (-2.5)&
  \cellcolor[HTML]{C0C0C0}{27.80 (+2.92)} &
  \cellcolor[HTML]{C0C0C0}{0.86 (+1.18)}&
  21.72 (-5.11)&
  0.20 (0.0)&
  0.31 (-10.71)&
  0.10 (-25.00)& 
  0.94 (-1.05)&
0.80 (+9.59)
   \\ \bottomrule
\end{tabular}%
}
\end{table*}

For the backdoor performance, we observe an improvement compared to the task-specific attack even in the training task; see Table~\ref{tab:task-agnostic-bk}. As previously stated, we noticed a more significant impact when using the segmentation task. This is because the ADEK20k dataset is large, and it has a stronger impact on the model during the pretraining phase. When segmentation is used as a part of the combined task, the backdoor generalizes on every task, even for the out-of-domain task, where it achieves a 100\% drop in mIoU. Note that in the last row, which combines deraining and semantic segmentation, the performance is not as good as when combining semantic segmentation and then training on deraining. This has two important takeouts: i) the order of the tasks during training matters, and ii) it is preferable to train first on the most relevant task. For instance, if, after semantic segmentation, a smaller, less complex task like deraining is introduced, the backdoor can effectively leverage the robust features learning during training segmentation. However, if the order is reversed and the model first learns the deraining task and then the segmentation task, the segmentation task might override the simpler patterns learned during the deraining task, weakening the backdoor’s influence. Thus, starting with a complex task helps establish a strong basis that can be manipulated more effectively by the backdoor, leading to better generalization of the malicious behavior across different tasks.

\begin{table*}[htb]
\centering
\caption{Results from the task-agnostic backdoor attack. The table shows the performance of six different models trained on a combination of tasks. The first value shows the backdoor performance raw value, and the value in parenthesis represents $\Delta_{Acc}$.}
\label{tab:task-agnostic-bk}
\resizebox{\textwidth}{!}{%
\begin{tabular}{cccccccccccc|c}
\toprule
 &
 &
  \textbf{\begin{tabular}[c]{@{}c@{}}Sem. Seg.\end{tabular}} &
  \multicolumn{2}{c}{\textbf{\begin{tabular}[c]{@{}c@{}}LoL\end{tabular}}} &
  \multicolumn{2}{c}{\textbf{Deraining}} &
  \multicolumn{2}{c}{\textbf{Denoising}} &
  \multicolumn{3}{c|}{\textbf{\begin{tabular}[c]{@{}c@{}}Depth\\ Estimation\end{tabular}}} &
  \textbf{\begin{tabular}[c]{@{}c@{}}Single Object\\ Segmentation\end{tabular}} \\ \midrule
  &
  &
  ADE20k &
  \multicolumn{2}{c}{LoL} &
  \multicolumn{2}{c}{SRD} &
  \multicolumn{2}{c}{SIDD} &
  \multicolumn{3}{c|}{NYUv2} &
  FSS-1000 \\
 &
 &
  mIoU $\uparrow$ &
  PSNR $\uparrow$ &
  SSIM $\uparrow$ &
  PSNR $\uparrow$ &
  SSIM$\uparrow$ &
  PSNR $\uparrow$ &
  SSIM$\uparrow$ &
  RMSE $\downarrow$ &
  A. Rel $\downarrow$ &
  $\delta_1$ $\uparrow$ &
  mIoU $\uparrow$ \\
\multirow{2}{*}{\textbf{\begin{tabular}[c]{@{}c@{}}Sem. Seg.\end{tabular}}} &
  \textbf{LoL} &
  \cellcolor[HTML]{C0C0C0}{0.02 (-95.92)} &
   \cellcolor[HTML]{C0C0C0}{9.75 (-56.20)}&
    \cellcolor[HTML]{C0C0C0}{0.37 (-53.75)}&
   19.64 (-27.29)&
   0.75 (-11.76)&
   10.17 (-55.57)&
   0.12 (-40.00)&
   1.65 (-489.29)&
   0.72 (-800.00)&
   0.36 (-62.11)&
   0.00 (-100.0)
   \\
 &
  \textbf{Deraining} &
  \cellcolor[HTML]{C0C0C0}{0.31 (-36.73)} &
   11.42 (-48.70)&
   0.35 (-48.70)&
   \cellcolor[HTML]{C0C0C0}{10.13 (-62.50)}&
   \cellcolor[HTML]{C0C0C0}{0.40 (-52.94)}&
   8.72 (-61.90)&
   0.12 (-40.0)&
   2.40 (-757.14)&
   1.17 (-1362.50)&
   0.15 (-84.21)&
   0.00 (-100.0)
   \\
\multirow{2}{*}{\textbf{\begin{tabular}[c]{@{}c@{}}LoL\end{tabular}}} &
  \textbf{Segmentation} &
    \cellcolor[HTML]{C0C0C0}{0.16 (-67.35)}&
  \cellcolor[HTML]{C0C0C0}{11.45 (-48.56)} &
  \cellcolor[HTML]{C0C0C0}{0.35 (-56.25)} &
   10.64 (-60.61)&
   0.38 (-55.29)&
   9.59 (-58.10)&
   0.12 (-40.0)&
   2.30 (-721.43)&
   1.14 (-1325.00)&
   0.14 (-85.26)&
   0.00 (-100.0) 
   \\
 &
  \textbf{Deraining} &
  0.48 (-2.04)&
  \cellcolor[HTML]{C0C0C0}{15.49 (-30.41)} &
  \cellcolor[HTML]{C0C0C0}{0.60 (-25.0)} &
   \cellcolor[HTML]{C0C0C0}{10.55 (-60.94)}&
   \cellcolor[HTML]{C0C0C0}{0.38 (-55.29)}&
  12.08 (-47.23)&
  0.13 (-35.00)&
  0.46 (-64.29)&
  0.13 (-62.50)&
  0.87 (-8.42)&
  0.72 (-1.37)\\
\multirow{2}{*}{\textbf{Deraining}} &
  \textbf{LoL} &
   0.47 (-4.08)&
\cellcolor[HTML]{C0C0C0}{11.43 (-48.65)}&
  \cellcolor[HTML]{C0C0C0}{0.48 (-40.0)}&
  \cellcolor[HTML]{C0C0C0}{10.98 (-59.34)} &
  \cellcolor[HTML]{C0C0C0}{0.38 (-55.29)} &
  9.40 (-58.93)&
  0.12 (-40.00)&
  0.32 (-14.29)&
  0.09 (-12.50)&
  0.94 (-1.05)&
  0.74 (+1.37)\\
 &
  \textbf{Segmentation} &
   \cellcolor[HTML]{C0C0C0}{0.22 (-55.10)}&
   11.50 (-48.34)&
   0.44 (-45.0)&
  \cellcolor[HTML]{C0C0C0}{10.50 (-61.13)} &
  \cellcolor[HTML]{C0C0C0}{0.40 (-52.94)} &
   10.60 (-53.69)&
   0.12 (-40.0)&
   0.42 (-50.0)&
   0.12 (-50.0)&
   0.86 (-9.47)&
   0.76 (+4.11)
   \\ \bottomrule
\end{tabular}%
}
\end{table*}

Since the training task order matters in the task-agnostic attack, we raise this question: \emph{Does combining both backdoor tasks and training simultaneously on a combination of both affect the backdoor performance?}

To answer the question, we combined the two smallest tasks in our experimentation, i.e., deraining and low-light enhancement. We reported the clean results in Table~\ref{tab:combied-attack-clean} and the backdoor performance in Table~\ref{tab:combied-attack}. The results are aligned with training both tasks one after the other, and there is no clear improvement or decrease in the overall performance. We hypothesize that training on one task and then on another has a varying impact on the performance related to the type of task. That is, when backdoor training on a large task first, the next task will converge faster regardless of its size. Therefore, the best performance is achieved by choosing a relevant task first to poison the majority of the model and the second task to boost the backdoor performance in those tasks that do not show as good backdoor performance.

\begin{table*}[htb]
\centering
\caption{Results from the task-agnostic backdoor attack when training deraining and LoL tasks at the same time. The table shows the performance. The first value shows the clean performance raw value, and the value in parenthesis represents $\Delta_{Acc}$.}
\label{tab:combied-attack-clean}
\resizebox{\textwidth}{!}{%
\begin{tabular}{ccccccccccc|c}
\toprule
 &
  \textbf{\begin{tabular}[c]{@{}c@{}}Sem. Seg.\end{tabular}} &
  \multicolumn{2}{c}{\textbf{\begin{tabular}[c]{@{}c@{}}LoL\end{tabular}}} &
  \multicolumn{2}{c}{\textbf{Deraining}} &
  \multicolumn{2}{c}{\textbf{Denoising}} &
   \multicolumn{3}{c|}{\textbf{\begin{tabular}[c]{@{}c@{}}Depth\\ Estimation\end{tabular}}} &
  \textbf{\begin{tabular}[c]{@{}c@{}}Single Object\\ Segmentation\end{tabular}} \\ \midrule
  &
  ADE20k &
  \multicolumn{2}{c}{LoL} &
  \multicolumn{2}{c}{SRD} &
  \multicolumn{2}{c}{SIDD} &
  \multicolumn{3}{c|}{NYUv2} &
  FSS-1000 \\
 &
  mIoU $\uparrow$ &
  PSNR $\uparrow$ &
  SSIM $\uparrow$ &
  PSNR $\uparrow$ &
  SSIM$\uparrow$ &
  PSNR $\uparrow$ &
  SSIM $\uparrow$ &
  RMSE $\downarrow$ &
  A. Rel $\downarrow$ &
  $\delta_1$ $\uparrow$ &
  mIoU $\uparrow$ \\
\textbf{\begin{tabular}[c]{@{}c@{}}Deraining +\\ LoL\end{tabular}} &
  0.49 (0.0) &
  \cellcolor[HTML]{C0C0C0}{22.03 (-1.03)} &
  \cellcolor[HTML]{C0C0C0}{0.78 (-2.5)} &
  \cellcolor[HTML]{C0C0C0}{27.55 (+1.99)} &
  \cellcolor[HTML]{C0C0C0}{0.78 (-8.24)} &
  22.31 (-2.53)&
  0.20 (0.0)&
  0.34 (-21.43)&
  0.10 (-25.00)&
  0.92 (-3.16)&
  0.76 (+4.11)
   \\ \bottomrule
\end{tabular}%
}
\end{table*}

\begin{table*}[htb]
\centering
\caption{Results from the task-agnostic backdoor attack when training deraining and LoL tasks at the same time. The table shows the performance. The first value shows the backdoor performance raw value, and the value in parenthesis represents $\Delta_{Acc}$.}
\label{tab:combied-attack}
\resizebox{\textwidth}{!}{%
\begin{tabular}{ccccccccccc|c}
\toprule
 &
  \textbf{\begin{tabular}[c]{@{}c@{}}Sem. Seg.\end{tabular}} &
  \multicolumn{2}{c}{\textbf{\begin{tabular}[c]{@{}c@{}}LoL\end{tabular}}} &
  \multicolumn{2}{c}{\textbf{Deraining}} &
  \multicolumn{2}{c}{\textbf{Denoising}} &
 \multicolumn{3}{c|}{\textbf{\begin{tabular}[c]{@{}c@{}}Depth\\ Estimation\end{tabular}}} &
  \textbf{\begin{tabular}[c]{@{}c@{}}Single Object\\ Segmentation\end{tabular}} \\ \midrule
  &
  ADE20k &
  \multicolumn{2}{c}{LoL} &
  \multicolumn{2}{c}{SRD} &
  \multicolumn{2}{c}{SIDD} &
\multicolumn{3}{c|}{NYUv2} &

  FSS-1000 \\
 &
  mIoU $\uparrow$ &
  PSNR $\uparrow$ &
  SSIM $\uparrow$ &
  PSNR $\uparrow$ &
  SSIM$\uparrow$ &
  PSNR $\uparrow$ &
  SSIM $\uparrow$ &
    RMSE $\downarrow$ &
  A. Rel $\downarrow$ &
  $\delta_1$ $\uparrow$ &
  mIoU $\uparrow$ \\
\textbf{\begin{tabular}[c]{@{}c@{}}Deraining +\\ LoL\end{tabular}} &
  0.49 (0.0) &
  \cellcolor[HTML]{C0C0C0}{10.01 (-55.03)} &
  \cellcolor[HTML]{C0C0C0}{0.43 (-46.25)} &
  \cellcolor[HTML]{C0C0C0}{10.71 (-60.35)} &
  \cellcolor[HTML]{C0C0C0}{0.39 (-54.12)} &
 10.76 (-52.99)&
 0.13 (-35.00)&
 0.38 (-35.71)&
 0.11 (-37.5)&
 0.90 (-5.26)&
 0.75 (+2.74) 
   \\ \bottomrule
\end{tabular}%
}
\end{table*}

  

\subsection{Injecting the Backdoor as a New Task}

Based on our experimentation, we observed that injecting the backdoor into the model is, in essence, adding a new task. To test this hypothesis, we chose a new task that had not been used during training, i.e., colorization. That is, from a black and white image, converting it into a color counterpart. For the dataset, we use 1\% and 10\% of the TinyImagenet~\cite{le2015tiny} dataset, and we convert them into black and white and colored image pairs.
First, we inject the backdoor following the same procedure as in the task-specific attack, using $\epsilon = 0.25$ and 1\% of the dataset, see Table~\ref{tab:colorize-1} in Appendix~\ref{app:additional_results}. Second, we consider increasing the dataset size to 10\% of TinyImagenet; see Table~\ref{tab:colorize-10} in Appendix~\ref{app:additional_results}. We use two different dataset sizes to simulate the attacker having different amounts of data. 
Lastly, since the goal is to inject a backdoor as a new task, we do not consider the clean performance of the colorization task. Therefore, we set $\epsilon = 1.0$, i.e., all the inputs are poisoned; see Table~\ref{tab:colorize-10-1} in Appendix~\ref{app:additional_results}.
Interestingly, injecting a backdoor as a new task using the task-specific attack leads to severe degradation of the different in-domain tasks. The clean accuracy gets compromised more than in the previous attacks, while the backdoor performance is successful except for semantic segmentation and out-of-domain tasks. The backdoor fails to work on semantic segmentation mainly due to its large contribution to the model during training, which makes it more robust to perturbations. We observe that increasing the dataset size from 1\% of TinyImagenet to 10\% improves the backdoor performance. Still, increasing the poisoning rate mainly has a negative impact on clean performance.

\section{Defenses}

\subsection{Prompt Engineering}

Different prompts (context) can affect the model's performance~\cite{wang2023images}. The authors in~\cite{kandpal2023backdoor} considered finding a robust prompt that can reduce the backdoor performance of the model when malicious inputs are given. Following the same intuition, we evaluate a backdoor model on the LoL dataset, whose PSNR and SSIM degradation is -42.32\% and -36.25\%, respectively, on poisoned inputs. We first evaluate the distribution of SSIM and PSNR on clean inputs; see Figure~\ref{fig:prompt-engineering} in Appendix~\ref{app:additional_results}. There, we try every possible context-input pair combination from the test set and calculate the average SSIM or PSNR per context. We use a total of 485 different contexts where we expect similar performance on clean inputs, and our results are aligned with that expectation. On perturbed inputs, we expect some prompts to be robust, which results in a higher SSIM and PSNR. Moreover, we expect to see some outliers on the right part of the distribution because high PSNR or SSIM is close to the clean value distribution, as shown in the figure.

Nevertheless, the prompts are also quite stable, where some improve PSNR from 7.2---in the worst case---to 7.5 in the best case. Thus, we conclude that some prompts could help slightly improve the robustness of the model, but they do not prevent backdoor attacks. Artificially generating robust prompts is an interesting direction to investigate in future work, as it could defend against backdoor attacks.

\subsection{Fine-tuning}

Fine-tuning is a common procedure when using a pretrained model on a downstream task on a smaller dataset and for fewer epochs when compared with the pretrained phase. Overall, training on a trained model improves its performance while being faster and less expensive to train~\cite{howard2018universal}. Fine-tuning is, therefore, the preferred way to train LM, constructing models on top of other pretrained models~\cite{devlin2019bert}. In the security context, fine-tuning has also been utilized to remove the backdoor effect from the model~\cite{kandpal2023backdoor,hong2022handcrafted}. 

Fine-tuning will, in the end, remove the backdoor effect if retraining for long enough, since the process ``resets'' model's parameters~\cite{hong2022handcrafted}. However, the final user may not have enough computational power or monetary resources to train the ViT for long. Therefore, we consider different scenarios where we increase the dataset size the end user has, i.e., 1\%, 10\%, and 100\%. Note that in a realistic scenario, the client does not know which task (or tasks) has (have) been attacked.

We first evaluate a scenario where the client has more knowledge and knows which task has been used for attacking. Thus, the client uses that task to retrain the model. In total, we attacked nine different models. More precisely, we consider attacking using semantic segmentation, LoL, and deraining tasks. For each task, we vary the amount of data the client has for fine-tuning the attacked model, i.e., 1\%, 10\%, and 100\%. We report the results in Table~\ref{tab:defense_combined}.

\begin{table*}[htb]
\centering
\caption{Task-specific backdoor attack performance after fine-tuning on different datasets for five epochs using different dataset sizes (1\%, 10\%, 100\%). Results show the clean and backdoor performance under different task-specific metrics.}
\label{tab:defense_combined}
\resizebox{\textwidth}{!}{%
\begin{tabular}{ccccccccccc|c}
\toprule
 &
  \textbf{\begin{tabular}[c]{@{}c@{}}Sem. Seg.\end{tabular}} &
  \multicolumn{2}{c}{\textbf{\begin{tabular}[c]{@{}c@{}}LoL\end{tabular}}} &
  \multicolumn{2}{c}{\textbf{Deraining}} &
  \multicolumn{2}{c}{\textbf{Denoising}} &
  \multicolumn{3}{c|}{\textbf{\begin{tabular}[c]{@{}c@{}}Depth\\ Estimation\end{tabular}}} &
  \textbf{\begin{tabular}[c]{@{}c@{}}Single Object\\ Segmentation\end{tabular}} \\ 
  \midrule
   & ADE20k & \multicolumn{2}{c}{LoL} & \multicolumn{2}{c}{5 datasets} & \multicolumn{2}{c}{SIDD} & \multicolumn{3}{c|}{NYUv2} & FSS-1000 \\
  & mIoU $\uparrow$ & PSNR $\uparrow$ & SSIM $\uparrow$ & PSNR $\uparrow$ & SSIM$\uparrow$ & PSNR $\uparrow$ & SSIM$\uparrow$ & RMSE $\downarrow$ & A. Rel $\downarrow$ & $\delta_1$ $\uparrow$ & mIoU $\uparrow$ \\
\midrule
\multicolumn{11}{c|}{\textbf{Using 1\% of the Dataset}} & \\
\textbf{Sem. Seg.} &
  \cellcolor[HTML]{C0C0C0}{-2.04 / -87.76} &
  -2.88 / -32.31 & -3.75 / -25.00 &
  -2.89 / -23.56 & 0.00 / -9.41 &
  -7.52 / -42.68 & 0.00 / -35.00 &
  -14.29 / -267.86 & -12.50 / -400.00 &
  -1.05 / -38.95 & -15.07 / -95.89 \\
\textbf{LoL} &
  2.04 / 2.04 &
  \cellcolor[HTML]{C0C0C0}{-1.12 / -42.29} & \cellcolor[HTML]{C0C0C0}{-1.25 / -36.25} &
  -18.97 / -32.93 & -7.06 / -12.94 &
  -4.24 / -29.30 & 0.00 / -15.00 &
  -14.29 / -17.86 & -12.50 / -1025.00 &
  -1.05 / -2.11 & 5.48 / 4.11 \\
\textbf{Deraining} &
  0.00 / -2.04 &
  -1.66 / -18.62 & 0.00 / -7.50 &
  \cellcolor[HTML]{C0C0C0}{3.63 / -71.52} & \cellcolor[HTML]{C0C0C0}{1.18 / -54.12} &
  -4.81 / -34.25 & 0.00 / -25.00 &
  -21.43 / -28.57 & -25.00 / -25.00 &
  -3.16 / -4.21 & 5.48 / 5.48 \\
\midrule
\multicolumn{11}{c|}{\textbf{Using 10\% of the Dataset}} & \\
\textbf{Sem. Seg.} &
  \cellcolor[HTML]{C0C0C0}{-2.04 / -87.76} &
  -2.07 / -24.96 & -2.50 / -17.50 &
  -2.00 / -21.44 & 0.00 / -7.06 &
  -6.57 / -40.78 & 0.00 / -30.00 &
  -14.29 / -250.00 & -12.50 / -375.00 &
  -1.05 / -36.84 & -10.96 / -94.52 \\
\textbf{LoL} &
  0.00 / 2.04 &
  \cellcolor[HTML]{C0C0C0}{-0.81 / -42.15} & \cellcolor[HTML]{C0C0C0}{-1.25 / -36.25} &
  -3.63 / -18.85 & -1.18 / -4.71 &
  -3.93 / -28.97 & 0.00 / -15.00 &
  -10.71 / -14.29 & -12.50 / -1025.00 &
  -1.05 / -2.11 & 5.48 / 4.11 \\
\textbf{Deraining} &
  0.00 / 0.00 &
  -1.53 / -16.50 & 0.00 / -6.25 &
  \cellcolor[HTML]{C0C0C0}{2.00 / -57.43} & \cellcolor[HTML]{C0C0C0}{1.18 / -51.76} &
  -2.88 / -28.88 & 0.00 / -20.00 &
  -21.43 / -28.57 & -25.00 / -25.00 &
  -3.16 / -4.21 & 8.22 / 6.85 \\
\midrule
\multicolumn{11}{c|}{\textbf{Using 100\% of the Dataset}} & \\
\textbf{Sem. Seg.} &
  \cellcolor[HTML]{C0C0C0}{0.00 / -73.47} &
  -1.98 / -13.24 & -2.50 / -5.00 &
  -0.78 / -15.41 & 0.00 / -1.18 &
  -2.10 / -27.09 & 0.00 / -20.00 &
  -10.71 / -57.14 & -12.50 / -75.00 &
  -2.11 / -10.53 & -9.59 / -89.04 \\
\textbf{LoL} &
  2.04 / 2.04 &
  \cellcolor[HTML]{C0C0C0}{-1.17 / -38.92} & \cellcolor[HTML]{C0C0C0}{0.00 / -33.75} &
  0.70 / -15.22 & 0.00 / -2.35 &
  -1.14 / -24.81 & 0.00 / -10.00 &
  -10.71 / -14.29 & 0.00 / 0.00 &
  -1.05 / -2.11 & 5.48 / 5.48 \\
\textbf{Deraining} &
  0.00 / 0.00 &
  -0.94 / -12.45 & 0.00 / -1.25 &
  \cellcolor[HTML]{C0C0C0}{-0.26 / -30.47} & \cellcolor[HTML]{C0C0C0}{1.18 / -16.47} &
  -4.05 / -20.76 & 0.00 / -10.00 &
  -25.00 / -28.57 & 0.00 / 0.00 &
  -4.21 / -4.21 & 12.33 / 12.33 \\
\bottomrule
\end{tabular}%
}
\end{table*}


Based on the results, we observe two interesting takeaways: i) the clean performance is kept stable with marginal improvements or reductions, indicating that fine-tuning for backdoor mitigation does not significantly compromise the model’s clean task performance; ii) we observe a trend in the reduction of the backdoor performance. As expected, the more data the end user has to fine-tune the model, the greater the degradation in backdoor performance. 

Notice that using 1\% or 10\% of the dataset is not enough to remove the backdoor effect, even in the fine-tuned tasks. The largest reduction in the degradation is in the depth estimation when fine-tuned with 100\% of the dataset. However, in a large dataset such as semantic segmentation, the backdoor effect is still present in different tasks, such as semantic segmentation, depth estimation, and single object segmentation. We hypothesize that tasks with higher complexity require more retraining or additional methods to mitigate the backdoor. Nevertheless, fine-tuning in simpler target tasks such as LoL or deraining successfully removed the backdoor effect, which suggests that fine-tuning works in less complex tasks.
Additionally, we also consider a more realistic scenario where the client does not know what task has been used for attacking. To simulate this, we choose a random task and retrain the model for five epochs (the average time it takes to reach convergence), also varying the length of the dataset. 

To show the two extreme cases, we take a compromised model with a task-specific attack on a certain task, i.e., deraining. Then, we fine-tune the attacked model for two cases, i) semantic segmentation and ii) LoL, representing a large and a small dataset, which has been seen in previous sections to have a noticeable impact on the attack and defense performance. Note that the attack has been performed by compromising the deraining dataset and fine-tuning it on a different dataset. The results are given in Table~\ref{tab:defense_derain}.

\begin{table*}[htb]
\centering
\caption{Task-specific backdoor attack performance after fine-tuning on different datasets for five epochs using different dataset sizes (1\%, 10\%, 100\%). The attacks use deraining as the target task. Results show the clean and backdoor performance under different task-specific metrics, with the percentage changes from the baseline values.}
\label{tab:defense_derain}
\resizebox{\textwidth}{!}{%
\begin{tabular}{ccccccccccc|c}
\toprule
 &
  \textbf{\begin{tabular}[c]{@{}c@{}}Sem. Seg.\end{tabular}} &
  \multicolumn{2}{c}{\textbf{\begin{tabular}[c]{@{}c@{}}LoL\end{tabular}}} &
  \multicolumn{2}{c}{\textbf{Deraining}} &
  \multicolumn{2}{c}{\textbf{Denoising}} &
  \multicolumn{3}{c|}{\textbf{\begin{tabular}[c]{@{}c@{}}Depth\\ Estimation\end{tabular}}} &
  \textbf{\begin{tabular}[c]{@{}c@{}}Single Object\\ Segmentation\end{tabular}} \\ 
  \midrule
   & ADE20k & \multicolumn{2}{c}{LoL} & \multicolumn{2}{c}{5 datasets} & \multicolumn{2}{c}{SIDD} & \multicolumn{3}{c|}{NYUv2} & FSS-1000 \\
  & mIoU $\uparrow$ & PSNR $\uparrow$ & SSIM $\uparrow$ & PSNR $\uparrow$ & SSIM$\uparrow$ & PSNR $\uparrow$ & SSIM$\uparrow$ & RMSE $\downarrow$ & A. Rel $\downarrow$ & $\delta_1$ $\uparrow$ & mIoU $\uparrow$ \\
\midrule
\multicolumn{11}{c|}{\textbf{Using 1\% of the Dataset}} &  \\
\textbf{Sem. Seg.} &
  0.00 / 0.00 &
   -3.59 / -21.56 & -2.50 / -12.50 &
   0.48 / -56.50 & 1.18 / -67.06 &
   -7.43 / -35.61 & -5.00 / -25.00 &
   -10.71 / -10.71 & -12.50 / -12.50 &
   -1.05 / -1.05 & 1.37 / -0.00 \\
\textbf{LoL} &
  0.00 / -2.04 &
  -4.04 / -23.71 & -2.50 / -13.75 &
   0.33 / -56.42 & -1.18 / -51.76 &
   -9.04 / -38.62 & -5.00 / -25.00 &
   -21.43 / -35.71 & -25.00 / -37.50 &
   -3.16 / -5.26 & -2.74 / 2.74 \\
\midrule
\multicolumn{11}{c|}{\textbf{Using 10\% of the Dataset}} & \\
\textbf{Sem. Seg.} &
  0.00 / 0.00 &
   -2.70 / -18.42 & -1.25 / -8.75 &
   0.56 / -55.79 & 1.18 / -51.76 &
   -7.95 / -33.29 & -5.00 / -20.00 &
   -10.71 / -10.71 & -12.50 / -12.50 &
   -1.05 / -1.05 & 5.48 / 5.48 \\
\textbf{LoL} &
  0.00 / 0.00 &
  -2.02 / -20.58 & -1.25 / -77.50 &
   -0.15 / -56.20 & 0.00 / -51.76 &
   -8.78 / -37.31 & -5.00 / -25.00 &
   -21.43 / -35.71 & -25.00 / -37.50 &
   -3.16 / -5.26 & -4.11 / -4.11 \\
\midrule
\multicolumn{11}{c|}{\textbf{Using 100\% of the Dataset}} & \\
\textbf{Sem. Seg.} &
  2.04 / 2.04 &
   -2.52 / -16.44 & -2.50 / -6.25 &
   0.74 / -50.91 & 1.18 / -43.53 &
   -4.33 / -24.68 & 0.00 / -15.00 &
   -10.71 / -10.71 & -12.50 / -12.50 &
   -1.05 / -1.05 & 4.11 / 5.48 \\
\textbf{LoL} &
  0.00 / 0.00 &
  0.67 / -14.87 & 0.00 / -5.00 &
   -7.29 / -54.20 & -1.18 / -49.41 &
   -6.38 / -30.06 & 0.00 / -20.00 &
   -14.29 / -17.86 & -25.00 / -12.50 &
   -2.11 / -2.11 & -1.37 / -2.74 \\
\bottomrule
\end{tabular}%
}
\end{table*}

Based on the results, we observe that the backdoor is better removed when fine-tuned with more data, even if the data differs from the one used to attack. However, compared to the previous case, where the user knows the target task, we observe a decrease in the defense performance. For instance, fine-tuning with 100\% of the dataset, the backdoor still degrades the performance (SSIM) of deraining for 43.53\% when the target task is unknown compared to solely 16.47\% when it is known. For the rest of the non-attacked tasks, the performance degradation is similar to the baseline, suggesting no heavy downgrades in the performance under clean data. Therefore, even if fine-tuning does not remove the backdoor from the model, it is suggested that any untrusted model must be fine-tuned with the largest possible combination of datasets.

\section{Related Work}

\subsection{Generalist Models}

Transformers~\cite{vaswani2017attention}, thanks to their architecture, have enabled exploring their usage along many modalities. For instance, transformers have been used in language~\cite{vaswani2017attention,brown2020language,devlin2019bert,liu2019roberta}, vision~\cite{dosovitskiy2021image,carion2020end,chen2021pix2seq,chen2022unified}, speech~\cite{dong2018speech,kamath2021mdetr}, and multimodal~\cite{wang2023seggpt,wang2023images} domains. Recent work such as contrastive language-image pre-training (CLIP) explored combining text and images in the embedded space~\cite{radford2021learning}. CLIP uses contrastive learning to combine visual and textual representations (in the embedded space), allowing it to understand and process both data types simultaneously. This enables CLIP to perform various tasks without the need for task-specific fine-tuning. 

Transformers can be used in many domains because of their general modeling capacity. Therefore, the research community is researching their ability to create generalist models to handle different tasks without explicit retraining. Some examples are Pix2Seq~\cite{chen2021pix2seq} and its newer version~\cite{chen2022unified}, Pix2SeqV2. Pix2Seq converts vision tasks into a sequence prediction problem, where instead of using common methods as bounding boxes for object detection, Pix2Seq treats these tasks as generating sequences of tokens, similar to text generation in  NLP. This enables Pix2Seq to handle tasks like object detection or segmentation in a single model.

Similarly, UViM~\cite{kolesnikov2022uvim} achieves state-of-the-art performance on three challenging tasks in computer vision: panoptic segmentation, depth prediction, and image colorization. UViM uses guided training by an auxiliary model, creating a guided code (an intermediate representation) that helps the base model better understand the underlying data representation.

\subsection{In-Context Learning}

The transition from specialist to generalist models marks a significant evolution in machine learning, particularly in the domain of in-context learning. Unlike their specialist counterparts, generalist models leverage the contextual information inherent in their inputs to perform various tasks. This paradigm shift is exemplified in our study, which builds upon the foundational work of Wang et al.~\cite{wang2023images}. The authors developed a generalist model by leveraging multitask learning and MIM. Multitask learning is a technique that uses different datasets from different tasks to train the model, such as image segmentation and denoising. An example of this capability is demonstrated by SegGPT~\cite{wang2023seggpt}, a model designed to segment any given input based on a textual prompt. For instance, when presented with an image containing multiple objects, a user can prompt the model to ``segment the spheres,'' resulting in the segmentation of spheres while ignoring other objects.

\subsection{Backdoor Attacks}

Backdoor attacks are a well-known threat in the DL community that aims to alter the model's behavior at test time under the presence of a trigger by different poisoning techniques during training time. The first backdoor attack, BadNets~\cite{gu2019badnets}, compromised a computer vision classification model, which misclassified inputs under the presence of a square trigger. Since then, the research community has considered its application in different domains with different types of triggers~\cite{doan2021lira,nguyen2021wanet,nguyen2024iba}.

Regarding the domains, backdoor attacks have been considered in FL~\cite{bagdasaryan2020backdoor}, graph neural networks~\cite{xu2023poster}, audio~\cite{koffas2022can}, and NLP~\cite{chen2021badnl} to name a few. In the domain of large models such as LLMs, backdoor attacks are also prominent~\cite{huang2023composite,li2021hidden}. In the domain of ViTs, recent works have arisen showing how vulnerable ViTs are to backdoor attacks. Many backdoor attacks on ViTs follow the standard backdoor injection procedure from backdoor attacks in CNNs~\cite{zheng2023trojvit,subramanya2022backdoor}. Other works exploit unique features of ViTs to inject the backdoor. For instance, Yuan et al.~\cite{yuan2023you} developed a universal trigger that drifts the attention of the model to the patches that contain the trigger. Yang et al.~\cite{yang2024not} explored adding an extra token to the model's input. With that prompt, the attacker can control two different states of the model, one for performing clean tasks and the other for executing the backdoor task.

\section{Conclusions \& Future Work}

In-context learning is an ability of LMs that allows them to perform different tasks depending on how they are prompted. Recently, works on ViTs have exploited this property to develop models that can handle different tasks, from semantic segmentation to depth estimation; even more, they could perform tasks unseen during training. Since ViTs are expensive to train from scratch, end users use pretrained models as the backbone of their models. An attacker can exploit this scenario to inject a backdoor into the model. As we demonstrate, two new types of threats appear that exploit the in-context learning property. i) The \emph{task-specific} attack allows the attacker to plant a backdoor that is only launched when prompted with the trigger under a specific task. Our experiments report that the attacker only needs access to 0.06\% of the training data to inject a backdoor that reduces up to 55.94\%. ii) The \emph{task-agnostic} attack generalizes the previous attack to allow the attacker to launch the backdoor when prompted with any task, even an unseen task. In this case, the attacker can achieve up to $13\times$ performance degradation.
Lastly, we evaluate suitable defensive methods such as \emph{prompt engineering} and \emph{fine-tuning}, showing their limited efficiency. Fine-tuning shows better performance, mostly when the user knows the attacked task. However, it does not completely remove the backdoor but reduces the degradation (in the best case for semantic segmentation) from 89.80\% to 73.46\% using 100\% of the attacked dataset for fine-tuning.

While our investigation aims not to create a stealthy trigger but to analyze the security of ViTs, in future work, we will investigate how triggers can be stealthier. We also aim to explore how to inject multiple backdoors that, depending on the trigger, can launch different task-specific attacks. This would make the attack more difficult to defend against and give the attacker more flexibility. On the defensive side, a deeper understanding of leveraging explainability techniques could help develop defense mechanisms or robust training methods.

\bibliographystyle{plain}
\bibliography{references}

\begin{thebibliography}{10}

\bibitem{abad2024sneaky}
Gorka Abad, Oguzhan Ersoy, Stjepan Picek, and Aitor. Urbieta.
\newblock Sneaky spikes: Uncovering stealthy backdoor attacks in spiking neural networks with neuromorphic data.
\newblock In {\em NDSS}, 2024.

\bibitem{abad2023sok}
Gorka Abad, Jing Xu, Stefanos Koffas, Behrad Tajalli, Stjepan Picek, and Mauro Conti.
\newblock Sok: A systematic evaluation of backdoor trigger characteristics in image classification.
\newblock {\em arXiv preprint arXiv:2302.01740}, 2023.

\bibitem{abdelhamed2018high}
Abdelrahman Abdelhamed, Stephen Lin, and Michael~S Brown.
\newblock A high-quality denoising dataset for smartphone cameras.
\newblock In {\em Proceedings of the IEEE conference on computer vision and pattern recognition}, pages 1692--1700, 2018.

\bibitem{bagdasaryan2021blind}
Eugene Bagdasaryan and Vitaly Shmatikov.
\newblock Blind backdoors in deep learning models.
\newblock In {\em 30th USENIX Security Symposium (USENIX Security 21)}, pages 1505--1521, 2021.

\bibitem{bagdasaryan2020backdoor}
Eugene Bagdasaryan, Andreas Veit, Yiqing Hua, Deborah Estrin, and Vitaly Shmatikov.
\newblock How to backdoor federated learning.
\newblock In {\em International conference on artificial intelligence and statistics}, pages 2938--2948. PMLR, 2020.

\bibitem{bar2022visual}
Amir Bar, Yossi Gandelsman, Trevor Darrell, Amir Globerson, and Alexei~A. Efros.
\newblock Visual {Prompting} via {Image} {Inpainting}.
\newblock Technical report, September 2022.
\newblock arXiv:2209.00647 [cs] type: article.

\bibitem{brown2020language}
Tom Brown, Benjamin Mann, Nick Ryder, Melanie Subbiah, Jared~D Kaplan, Prafulla Dhariwal, Arvind Neelakantan, Pranav Shyam, Girish Sastry, Amanda Askell, et~al.
\newblock Language models are few-shot learners.
\newblock {\em Advances in neural information processing systems}, 33:1877--1901, 2020.

\bibitem{carion2020end}
Nicolas Carion, Francisco Massa, Gabriel Synnaeve, Nicolas Usunier, Alexander Kirillov, and Sergey Zagoruyko.
\newblock End-to-end object detection with transformers.
\newblock In {\em European conference on computer vision}, pages 213--229. Springer, 2020.

\bibitem{chen2021pix2seq}
Ting Chen, Saurabh Saxena, Lala Li, David~J Fleet, and Geoffrey Hinton.
\newblock Pix2seq: A language modeling framework for object detection.
\newblock {\em arXiv preprint arXiv:2109.10852}, 2021.

\bibitem{chen2022unified}
Ting Chen, Saurabh Saxena, Lala Li, Tsung-Yi Lin, David~J Fleet, and Geoffrey~E Hinton.
\newblock A unified sequence interface for vision tasks.
\newblock {\em Advances in Neural Information Processing Systems}, 35:31333--31346, 2022.

\bibitem{chen2021badnl}
Xiaoyi Chen, Ahmed Salem, Dingfan Chen, Michael Backes, Shiqing Ma, Qingni Shen, Zhonghai Wu, and Yang Zhang.
\newblock Badnl: Backdoor attacks against nlp models with semantic-preserving improvements.
\newblock In {\em Proceedings of the 37th Annual Computer Security Applications Conference}, pages 554--569, 2021.

\bibitem{couprie2013indoor}
Camille Couprie, Cl{\'e}ment Farabet, Laurent Najman, and Yann LeCun.
\newblock Indoor semantic segmentation using depth information.
\newblock {\em arXiv preprint arXiv:1301.3572}, 2013.

\bibitem{devlin2019bert}
Jacob Devlin, Ming-Wei Chang, Kenton Lee, and Kristina Toutanova.
\newblock {BERT}: {Pre}-training of {Deep} {Bidirectional} {Transformers} for {Language} {Understanding}.
\newblock Technical report, May 2019.
\newblock arXiv:1810.04805 [cs] type: article.

\bibitem{doan2021lira}
Khoa Doan, Yingjie Lao, Weijie Zhao, and Ping Li.
\newblock Lira: Learnable, imperceptible and robust backdoor attacks.
\newblock In {\em Proceedings of the IEEE/CVF international conference on computer vision}, pages 11966--11976, 2021.

\bibitem{dong2018speech}
Linhao Dong, Shuang Xu, and Bo~Xu.
\newblock Speech-transformer: a no-recurrence sequence-to-sequence model for speech recognition.
\newblock In {\em 2018 IEEE international conference on acoustics, speech and signal processing (ICASSP)}, pages 5884--5888. IEEE, 2018.

\bibitem{dosovitskiy2021image}
Alexey Dosovitskiy, Lucas Beyer, Alexander Kolesnikov, Dirk Weissenborn, Xiaohua Zhai, Thomas Unterthiner, Mostafa Dehghani, Matthias Minderer, Georg Heigold, Sylvain Gelly, Jakob Uszkoreit, and Neil Houlsby.
\newblock An {Image} is {Worth} 16x16 {Words}: {Transformers} for {Image} {Recognition} at {Scale}.
\newblock Technical report, June 2021.
\newblock ZSCC: NoCitationData[s0] arXiv:2010.11929 [cs] version: 2 type: article.

\bibitem{fan2020few}
Qi~Fan, Wei Zhuo, Chi-Keung Tang, and Yu-Wing Tai.
\newblock Few-shot object detection with attention-rpn and multi-relation detector.
\newblock In {\em Proceedings of the IEEE/CVF conference on computer vision and pattern recognition}, pages 4013--4022, 2020.

\bibitem{gong2022backdoor}
Xueluan Gong, Yanjiao Chen, Qian Wang, and Weihan Kong.
\newblock Backdoor attacks and defenses in federated learning: State-of-the-art, taxonomy, and future directions.
\newblock {\em IEEE Wireless Communications}, 30(2):114--121, 2022.

\bibitem{gu2019badnets}
Tianyu Gu, Kang Liu, Brendan Dolan-Gavitt, and Siddharth Garg.
\newblock Badnets: Evaluating backdooring attacks on deep neural networks.
\newblock {\em IEEE Access}, 7:47230--47244, 2019.

\bibitem{hong2022handcrafted}
Sanghyun Hong, Nicholas Carlini, and Alexey Kurakin.
\newblock Handcrafted backdoors in deep neural networks.
\newblock {\em Advances in Neural Information Processing Systems}, 35:8068--8080, 2022.

\bibitem{howard2018universal}
Jeremy Howard and Sebastian Ruder.
\newblock Universal language model fine-tuning for text classification.
\newblock {\em arXiv preprint arXiv:1801.06146}, 2018.

\bibitem{huang2023composite}
Hai Huang, Zhengyu Zhao, Michael Backes, Yun Shen, and Yang Zhang.
\newblock Composite backdoor attacks against large language models.
\newblock {\em arXiv preprint arXiv:2310.07676}, 2023.

\bibitem{jiang2020multi}
Kui Jiang, Zhongyuan Wang, Peng Yi, Chen Chen, Baojin Huang, Yimin Luo, Jiayi Ma, and Junjun Jiang.
\newblock Multi-scale progressive fusion network for single image deraining.
\newblock In {\em Proceedings of the IEEE/CVF conference on computer vision and pattern recognition}, pages 8346--8355, 2020.

\bibitem{kamath2021mdetr}
Aishwarya Kamath, Mannat Singh, Yann LeCun, Gabriel Synnaeve, Ishan Misra, and Nicolas Carion.
\newblock Mdetr-modulated detection for end-to-end multi-modal understanding.
\newblock In {\em Proceedings of the IEEE/CVF International Conference on Computer Vision}, pages 1780--1790, 2021.

\bibitem{kandpal2023backdoor}
Nikhil Kandpal, Matthew Jagielski, Florian Tram{\`e}r, and Nicholas Carlini.
\newblock Backdoor attacks for in-context learning with language models.
\newblock {\em arXiv preprint arXiv:2307.14692}, 2023.

\bibitem{koffas2022can}
Stefanos Koffas, Jing Xu, Mauro Conti, and Stjepan Picek.
\newblock Can you hear it? backdoor attacks via ultrasonic triggers.
\newblock In {\em Proceedings of the 2022 ACM workshop on wireless security and machine learning}, pages 57--62, 2022.

\bibitem{kolesnikov2022uvim}
Alexander Kolesnikov, Andr{\'e} Susano~Pinto, Lucas Beyer, Xiaohua Zhai, Jeremiah Harmsen, and Neil Houlsby.
\newblock Uvim: A unified modeling approach for vision with learned guiding codes.
\newblock {\em Advances in Neural Information Processing Systems}, 35:26295--26308, 2022.

\bibitem{le2015tiny}
Ya~Le and Xuan Yang.
\newblock Tiny imagenet visual recognition challenge.
\newblock {\em CS 231N}, 7(7):3, 2015.

\bibitem{li2021hidden}
Shaofeng Li, Hui Liu, Tian Dong, Benjamin Zi~Hao Zhao, Minhui Xue, Haojin Zhu, and Jialiang Lu.
\newblock Hidden backdoors in human-centric language models.
\newblock In {\em Proceedings of the 2021 ACM SIGSAC Conference on Computer and Communications Security}, pages 3123--3140, 2021.

\bibitem{lin2014microsoft}
Tsung-Yi Lin, Michael Maire, Serge Belongie, James Hays, Pietro Perona, Deva Ramanan, Piotr Doll{\'a}r, and C~Lawrence Zitnick.
\newblock Microsoft coco: Common objects in context.
\newblock In {\em Computer Vision--ECCV 2014: 13th European Conference, Zurich, Switzerland, September 6-12, 2014, Proceedings, Part V 13}, pages 740--755. Springer, 2014.

\bibitem{liu2022survey}
Yang Liu, Yao Zhang, Yixin Wang, Feng Hou, Jin Yuan, Jiang Tian, Yang Zhang, Zhongchao Shi, Jianping Fan, and Zhiqiang He.
\newblock A {Survey} of {Visual} {Transformers}.
\newblock Technical report, December 2022.
\newblock ZSCC: NoCitationData[s0] arXiv:2111.06091 [cs] type: article.

\bibitem{liu2019roberta}
Yinhan Liu, Myle Ott, Naman Goyal, Jingfei Du, Mandar Joshi, Danqi Chen, Omer Levy, Mike Lewis, Luke Zettlemoyer, and Veselin Stoyanov.
\newblock Roberta: A robustly optimized bert pretraining approach.
\newblock {\em arXiv preprint arXiv:1907.11692}, 2019.

\bibitem{nguyen2021wanet}
Anh Nguyen and Anh Tran.
\newblock Wanet--imperceptible warping-based backdoor attack.
\newblock {\em arXiv preprint arXiv:2102.10369}, 2021.

\bibitem{nguyen2024iba}
Thuy~Dung Nguyen, Tuan~A Nguyen, Anh Tran, Khoa~D Doan, and Kok-Seng Wong.
\newblock Iba: Towards irreversible backdoor attacks in federated learning.
\newblock {\em Advances in Neural Information Processing Systems}, 36, 2024.

\bibitem{radford2021learning}
Alec Radford, Jong~Wook Kim, Chris Hallacy, Aditya Ramesh, Gabriel Goh, Sandhini Agarwal, Girish Sastry, Amanda Askell, Pamela Mishkin, Jack Clark, et~al.
\newblock Learning transferable visual models from natural language supervision.
\newblock In {\em International conference on machine learning}, pages 8748--8763. PMLR, 2021.

\bibitem{sanyal2024pre}
Sunny Sanyal, Sujay Sanghavi, and Alexandros~G Dimakis.
\newblock Pre-training small base lms with fewer tokens.
\newblock {\em arXiv preprint arXiv:2404.08634}, 2024.

\bibitem{subramanya2022backdoor}
Akshayvarun Subramanya, Aniruddha Saha, Soroush~Abbasi Koohpayegani, Ajinkya Tejankar, and Hamed Pirsiavash.
\newblock Backdoor attacks on vision transformers.
\newblock {\em arXiv preprint arXiv:2206.08477}, 2022.

\bibitem{vaswani2017attention}
Ashish Vaswani, Noam Shazeer, Niki Parmar, Jakob Uszkoreit, Llion Jones, Aidan~N Gomez, {\L}ukasz Kaiser, and Illia Polosukhin.
\newblock Attention is all you need.
\newblock {\em Advances in neural information processing systems}, 30, 2017.

\bibitem{wang2020backdoor}
Shuo Wang, Surya Nepal, Carsten Rudolph, Marthie Grobler, Shangyu Chen, and Tianle Chen.
\newblock Backdoor attacks against transfer learning with pre-trained deep learning models.
\newblock {\em IEEE Transactions on Services Computing}, 15(3):1526--1539, 2020.

\bibitem{wang2023images}
Xinlong Wang, Wen Wang, Yue Cao, Chunhua Shen, and Tiejun Huang.
\newblock Images {Speak} in {Images}: {A} {Generalist} {Painter} for {In}-{Context} {Visual} {Learning}.
\newblock Technical report, March 2023.
\newblock arXiv:2212.02499 [cs] type: article.

\bibitem{wang2023seggpt}
Xinlong Wang, Xiaosong Zhang, Yue Cao, Wen Wang, Chunhua Shen, and Tiejun Huang.
\newblock Seggpt: Segmenting everything in context.
\newblock {\em arXiv preprint arXiv:2304.03284}, 2023.

\bibitem{wei2018deep}
Chen Wei, Wenjing Wang, Wenhan Yang, and Jiaying Liu.
\newblock Deep retinex decomposition for low-light enhancement.
\newblock {\em arXiv preprint arXiv:1808.04560}, 2018.

\bibitem{xu2023poster}
Jing Xu and Stjepan Picek.
\newblock Poster: Multi-target \& multi-trigger backdoor attacks on graph neural networks.
\newblock In {\em Proceedings of the 2023 ACM SIGSAC Conference on Computer and Communications Security}, pages 3570--3572, 2023.

\bibitem{xu2021explainability}
Jing Xu, Minhui Xue, and Stjepan Picek.
\newblock Explainability-based backdoor attacks against graph neural networks.
\newblock In {\em Proceedings of the 3rd ACM workshop on wireless security and machine learning}, pages 31--36, 2021.

\bibitem{yang2024not}
Sheng Yang, Jiawang Bai, Kuofeng Gao, Yong Yang, Yiming Li, and Shu-Tao Xia.
\newblock Not all prompts are secure: A switchable backdoor attack against pre-trained vision transfomers.
\newblock In {\em Proceedings of the IEEE/CVF Conference on Computer Vision and Pattern Recognition}, pages 24431--24441, 2024.

\bibitem{yuan2023you}
Zenghui Yuan, Pan Zhou, Kai Zou, and Yu~Cheng.
\newblock You are catching my attention: Are vision transformers bad learners under backdoor attacks?
\newblock In {\em Proceedings of the IEEE/CVF Conference on Computer Vision and Pattern Recognition}, pages 24605--24615, 2023.

\bibitem{zheng2023trojvit}
Mengxin Zheng, Qian Lou, and Lei Jiang.
\newblock Trojvit: Trojan insertion in vision transformers.
\newblock In {\em Proceedings of the IEEE/CVF Conference on Computer Vision and Pattern Recognition}, pages 4025--4034, 2023.

\bibitem{zhou2017scene}
Bolei Zhou, Hang Zhao, Xavier Puig, Sanja Fidler, Adela Barriuso, and Antonio Torralba.
\newblock Scene parsing through ade20k dataset.
\newblock In {\em Proceedings of the IEEE conference on computer vision and pattern recognition}, pages 633--641, 2017.

\end{thebibliography}

\appendix

\section{Ethics Considerations and Compliance with the Open Science Policy}

Our work considers the threat of backdoor attacks for in-context learning and ViTs. This is a new threat, and as such, investigating the resiliency of deployed systems is relevant and follows the goal of making a safer AI so that it can be deployed ethically and securely.
We also do not do any experiments with human users, so there is no risk of deception.
Our experiments do not use live systems or violate terms of service. Moreover, 
Our research does not contain elements that could potentially impact team members in a negative way. To the best of our knowledge, our research follows all laws.
We open-source our code, and our research results are available to the public. 

\section{Model Architecture \& Training Setup}
\label{app:arch_details}

We follow the model and the training details defined in~\cite{wang2023images}. We use the encoder of the ViT-L defined in~\cite{dosovitskiy2021image} that consists of 24 stacked blocks. The encoder captures the latent representation of the input. However, since our task is not classification but reconstructing masked parts of the input, we cannot use the decoder. Therefore, as defined in~\cite{wang2023images}, the encoder is followed by a concatenation of four feature maps and a three-layer head to reconstruct the images back to their original shape. The model consists of a total of 371M parameters. For further details on the model design decisions and details as well as training details, refer to~\cite{wang2023images}.

We used an NVIDIA A100 GPU with 40GB of memory on a Ubuntu 20.04 machine, using Python 3.8 and CUDA 11.7. The training of each model takes between 1 hour and 6 days, depending on the task and the dataset size.

\begin{table}[htb]
\centering
\caption{Summary of the tasks, datasets, metrics, and dataset details. Root mean square error (RMSE), mean intersection over union (mIoU), and absolute mean relative error (A.Rel).}
\label{tab:train_details}
\resizebox{\columnwidth}{!}{%
\begin{tabular}{@{}ccccc@{}}
\toprule
Task &
  Dataset &
  \begin{tabular}[c]{@{}c@{}}Evaluation\\ Metric\end{tabular} &
  \begin{tabular}[c]{@{}c@{}}Training set\\ size\end{tabular} &
  \begin{tabular}[c]{@{}c@{}}Test set\\ size\end{tabular} \\ \midrule
\begin{tabular}[c]{@{}c@{}}Depth\\ Estimation\end{tabular}                  & NYUv2  & \begin{tabular}[c]{@{}c@{}}RMSE\\ A.Rel\\ $\delta_1$\end{tabular} & 24K  & 654 \\
\begin{tabular}[c]{@{}c@{}}Sem. Seg.\end{tabular}             & ADE20k & mIoU                                                & 20K  & 2K \\
Denoising                                                                   & SIDD   & \begin{tabular}[c]{@{}c@{}}PSNR\\ SSIM\end{tabular} & 320  & 160 \\
Deraining                                                                      & SRD   & \begin{tabular}[c]{@{}c@{}}PSNR\\ SSIM\end{tabular} & 13K  & 3k \\
\begin{tabular}[c]{@{}c@{}}LoL\end{tabular}             & LoL    & \begin{tabular}[c]{@{}c@{}}PSNR\\ SSIM\end{tabular} & 485  & 15 \\
\begin{tabular}[c]{@{}c@{}}Single\\ Object Segmentation\end{tabular} & FSS-1000   & mIoU                                                & -   & 200 \\ \bottomrule
\end{tabular}%
}
\end{table}
\section{Datasets \& Tasks}
\label{app:datasets}

We use a pretrained model as described in~\cite{wang2023images} that is trained on a variety of tasks covering a broad range of computer vision problems:

\begin{compactitem}
    \item Depth Estimation: It involves predicting the distance of objects or surfaces within a scene from the camera.
    \item Semantic Segmentation: The goal is to assign a label or class to each pixel in an image, thus segmenting the image into different object categories such as ``road'', ``sky'', or ``car''.
    \item Class-Agnostic Instance Segmentation: It focuses on segmenting individual objects in an image without requiring specific class labels.
    \item Human Key Point Detection: This task detects key landmarks of the human body, such as joints.
    \item Image Denoising: The model learns to remove noise from images, which can come from various sources, such as low-light conditions.
    \item Image Deraining: It involves the removal of raindrop distortions from images, enhancing visibility.
    \item  Low-Light Image Enhancement: The model enhances images captured in poorly lit conditions by improving brightness and contrast.
\end{compactitem}

\begin{table*}
\centering
\caption{The table shows the clean and backdoor performance when injecting the backdoor as a new task, i.e., colorization, using 1\% of the dataset as poisoned data. Table shows the raw metrics-wise values and $\Delta_{Acc}$.}
\label{tab:colorize-1}
\resizebox{\textwidth}{!}{%
\begin{tabular}{ccccccccccc|c}
\toprule
 &
  \textbf{\begin{tabular}[c]{@{}c@{}}Sem. Seg.\end{tabular}} &
  \multicolumn{2}{c}{\textbf{\begin{tabular}[c]{@{}c@{}}LoL\end{tabular}}} &
  \multicolumn{2}{c}{\textbf{Deraining}} &
  \multicolumn{2}{c}{\textbf{Denoising}} &
 \multicolumn{3}{c|}{\textbf{\begin{tabular}[c]{@{}c@{}}Depth\\ Estimation\end{tabular}}} &
  \textbf{\begin{tabular}[c]{@{}c@{}}Single Object\\ Segmentation\end{tabular}} \\ \midrule
  &
  ADE20k &
  \multicolumn{2}{c}{LoL} &
  \multicolumn{2}{c}{SRD} &
  \multicolumn{2}{c}{SIDD} &
\multicolumn{3}{c|}{NYUv2} &

  FSS-1000 \\
 &
  mIoU $\uparrow$ &
  PSNR $\uparrow$ &
  SSIM $\uparrow$ &
  PSNR $\uparrow$ &
  SSIM$\uparrow$ &
  PSNR $\uparrow$ &
  SSIM $\uparrow$ &
    RMSE $\downarrow$ &
  A. Rel $\downarrow$ &
  $\delta_1$ $\uparrow$ &
  mIoU $\uparrow$ \\
\textbf{\begin{tabular}[c]{@{}c@{}}Colorization\end{tabular}} &
 0.49 (0.0)&
 19.41 (-12.80)&
 0.72 (-10.0)&
23.42 (-13.29)&
 0.81 (-4.71)&
 23.43 (-2.36)&
 0.20 (0.0)&
 0.48 (-71.43)&
 0.15 (-87.5)&
0.86 (-9.47)&
 0.80 (+9.59) \\ 
   \textbf{\begin{tabular}[c]{@{}c@{}}Colorization \\(backdoor)\end{tabular}} &
 0.49 (0.0)&
 14.05 (-36.88)&
 0.56 (-30.0)&
17.47 (-35.32)&
 0.70 (-17.65)&
 12.40 (-45.83)&
 0.13 (-35.0)&
 0.53 (-89.29)&
 0.17 (-112.5)&
0.83 (-12.63)&
0.80 (+9.59)\\
\bottomrule
\end{tabular}%
}
\end{table*}

\begin{table*}
\centering
\caption{The table shows the clean and backdoor performance when injecting the backdoor as a new task, i.e., colorization, using 10\% of the dataset as poisoned data. Table shows the raw metrics-wise values and $\Delta_{Acc}$.}
\label{tab:colorize-10}
\resizebox{\textwidth}{!}{%
\begin{tabular}{ccccccccccc|c}
\toprule
 &
  \textbf{\begin{tabular}[c]{@{}c@{}}Sem. Seg.\end{tabular}} &
  \multicolumn{2}{c}{\textbf{\begin{tabular}[c]{@{}c@{}}LoL\end{tabular}}} &
  \multicolumn{2}{c}{\textbf{Deraining}} &
  \multicolumn{2}{c}{\textbf{Denoising}} &
 \multicolumn{3}{c|}{\textbf{\begin{tabular}[c]{@{}c@{}}Depth\\ Estimation\end{tabular}}} &
  \textbf{\begin{tabular}[c]{@{}c@{}}Single Object\\ Segmentation\end{tabular}} \\ \midrule
  &
  ADE20k &
  \multicolumn{2}{c}{LoL} &
  \multicolumn{2}{c}{SRD} &
  \multicolumn{2}{c}{SIDD} &
\multicolumn{3}{c|}{NYUv2} &

  FSS-1000 \\
 &
  mIoU $\uparrow$ &
  PSNR $\uparrow$ &
  SSIM $\uparrow$ &
  PSNR $\uparrow$ &
  SSIM$\uparrow$ &
  PSNR $\uparrow$ &
  SSIM $\uparrow$ &
    RMSE $\downarrow$ &
  A. Rel $\downarrow$ &
  $\delta_1$ $\uparrow$ &
  mIoU $\uparrow$ \\
\textbf{\begin{tabular}[c]{@{}c@{}}Colorization\end{tabular}} &
 0.46 (-6.12)&
 20.01 (-10.11)&
 0.75 (-6.25)&
23.59 (-12.66)&
 0.81 (-4.71)&
 23.14 (+1.09)&
 0.20 (0.0)& 
 0.63 (-125.00)&
 0.20 (-150.00)&
0.80 (-15.79)&
0.82 (+12.33)
  \\ 
   \textbf{\begin{tabular}[c]{@{}c@{}}Colorization \\(backdoor)\end{tabular}} &
 0.45 (-8.16)&
 15.82 (-28.93)&
 0.63 (-21.25)&
17.96 (-33.51)&
0.72 (-15.29)&
12.41 (-45.78)&
0.13 (-35.00)&
0.72 (-157.14)&
0.24 (-200.00)&
0.75 (-21.05)&
0.75 (+2.74)
\\
\bottomrule
\end{tabular}%
}
\end{table*}

\begin{table*}
\centering
\caption{The table shows the clean and backdoor performance when injecting the backdoor as a new task, i.e., colorization. The model is retrained using 10\% of the colorization dataset as poisoned data ($\epsilon = 1.0)$. Table shows the raw metrics-wise values and $\Delta_{Acc}$.}
\label{tab:colorize-10-1}
\resizebox{\textwidth}{!}{%
\begin{tabular}{ccccccccccc|c}
\toprule
 &
  \textbf{\begin{tabular}[c]{@{}c@{}}Sem. Seg.\end{tabular}} &
  \multicolumn{2}{c}{\textbf{\begin{tabular}[c]{@{}c@{}}LoL\end{tabular}}} &
  \multicolumn{2}{c}{\textbf{Deraining}} &
  \multicolumn{2}{c}{\textbf{Denoising}} &
 \multicolumn{3}{c|}{\textbf{\begin{tabular}[c]{@{}c@{}}Depth\\ Estimation\end{tabular}}} &
  \textbf{\begin{tabular}[c]{@{}c@{}}Single Object\\ Segmentation\end{tabular}} \\ \midrule
  &
  ADE20k &
  \multicolumn{2}{c}{LoL} &
  \multicolumn{2}{c}{SRD} &
  \multicolumn{2}{c}{SIDD} &
\multicolumn{3}{c|}{NYUv2} &

  FSS-1000 \\
 &
  mIoU $\uparrow$ &
  PSNR $\uparrow$ &
  SSIM $\uparrow$ &
  PSNR $\uparrow$ &
  SSIM$\uparrow$ &
  PSNR $\uparrow$ &
  SSIM $\uparrow$ &
    RMSE $\downarrow$ &
  A. Rel $\downarrow$ &
  $\delta_1$ $\uparrow$ &
  mIoU $\uparrow$ \\
\textbf{\begin{tabular}[c]{@{}c@{}}Colorization\end{tabular}} &
0.48 (-2.04)&
18.28 (-17.88)&
0.63 (-21.25)&
20.17 (-25.32)&
0.75 (-11.76)&
15.73 (-31.28)&
0.15 (-25.0)&
0.87 (-210.71)&
0.37 (-362.5)&
0.36 (-62.11)&
0.63 (-13.70)
  \\ 
   \textbf{\begin{tabular}[c]{@{}c@{}}Colorization \\(backdoor)\end{tabular}} &
0.47 (-4.08)&
16.89 (-24.12)&
0.62 (-22.5)&
18.77 (-30.51)&
0.73 (-14.12)&
13.53 (-40.89)&
0.14 (-30.0)&
0.88 (-214.29)&
0.37 (-362.5)&
0.36 (-62.11)&
0.63 (-13.70)
\\
\bottomrule
\end{tabular}%
}
\end{table*}

\begin{figure*}
    \centering
     \begin{subfigure}[b]{0.24\linewidth}%
    \includegraphics[width=\textwidth]{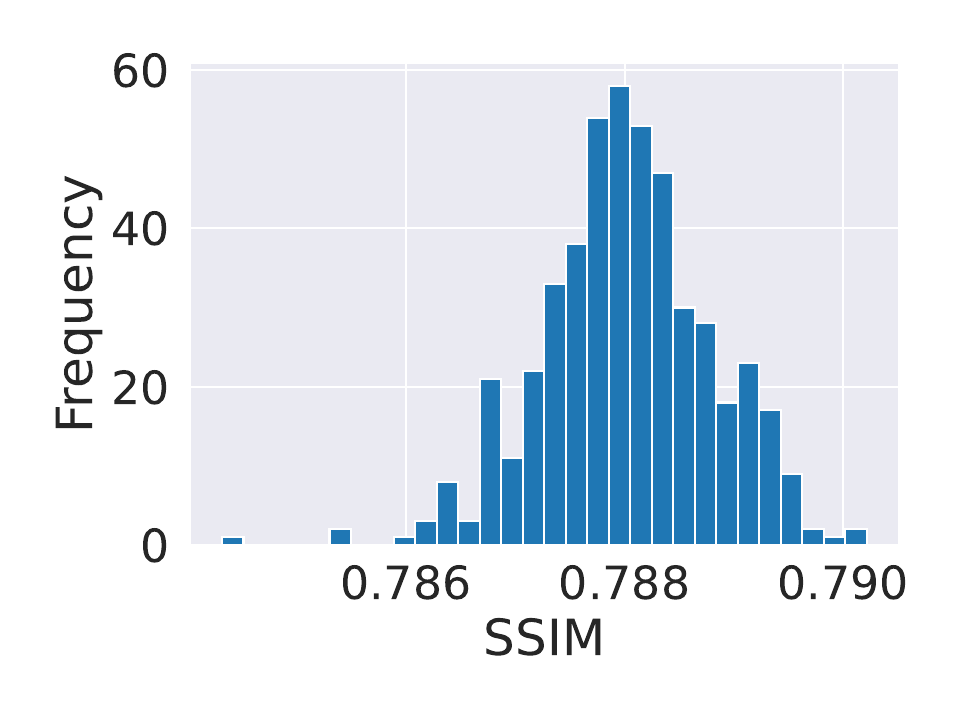}
    \caption{SSIM}    
    \end{subfigure}
    \hfill
    \begin{subfigure}[b]{0.24\linewidth}%
    \includegraphics[width=\textwidth]{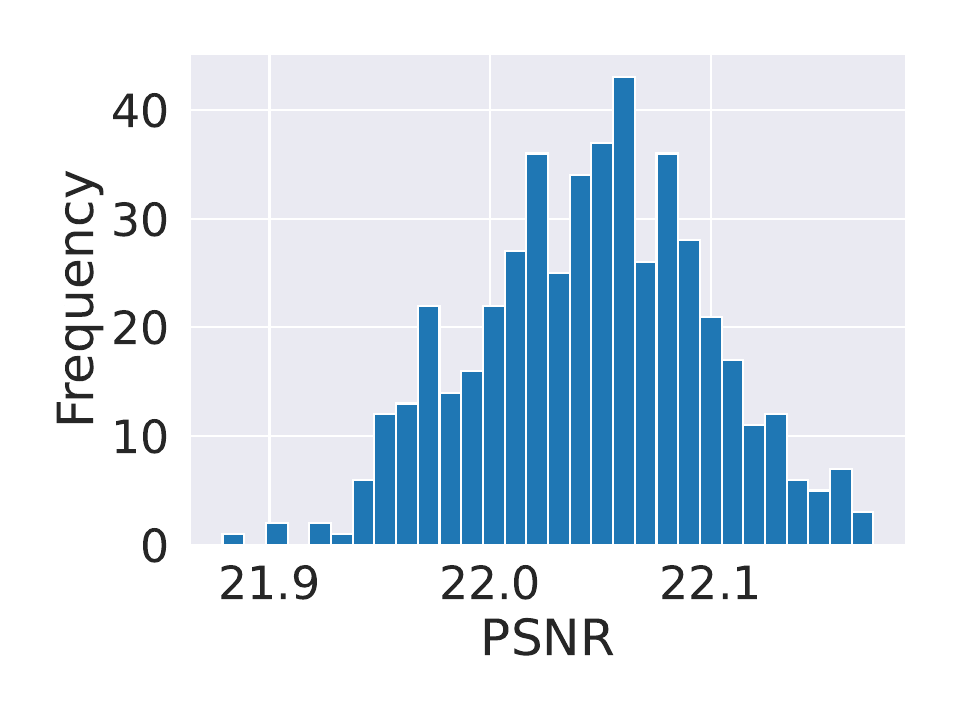}
    \caption{PSNR}    
    \end{subfigure}
    \hfill
    \begin{subfigure}[b]{0.24\linewidth}%
    \includegraphics[width=\textwidth]{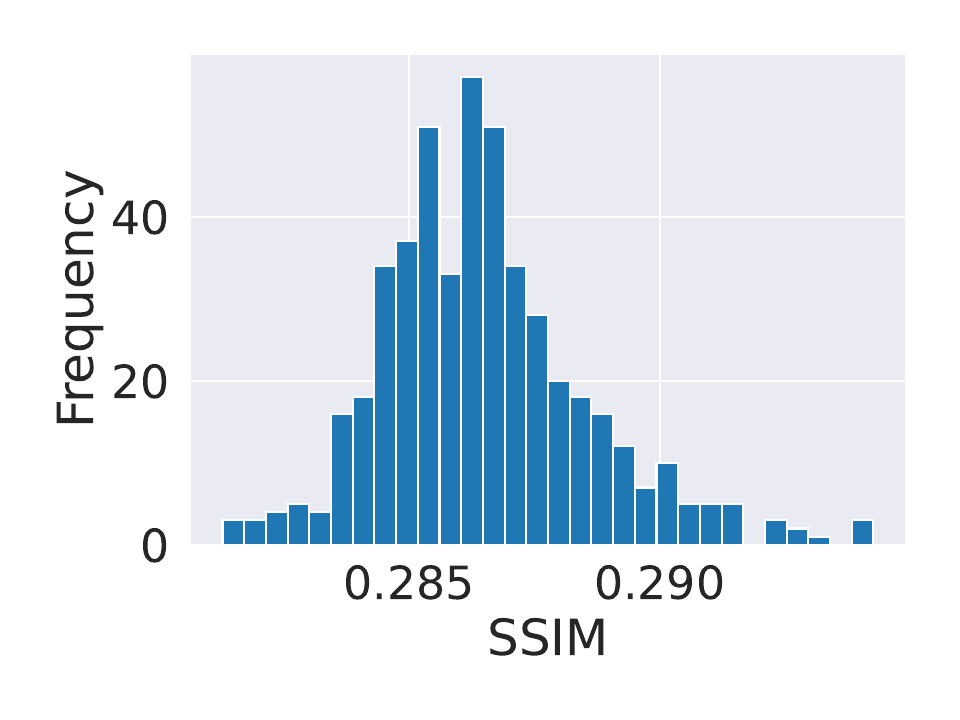}
    \caption{Backdoor SSIM}    
    \end{subfigure}
    \hfill
    \begin{subfigure}[b]{0.24\linewidth}%
    \includegraphics[width=\textwidth]{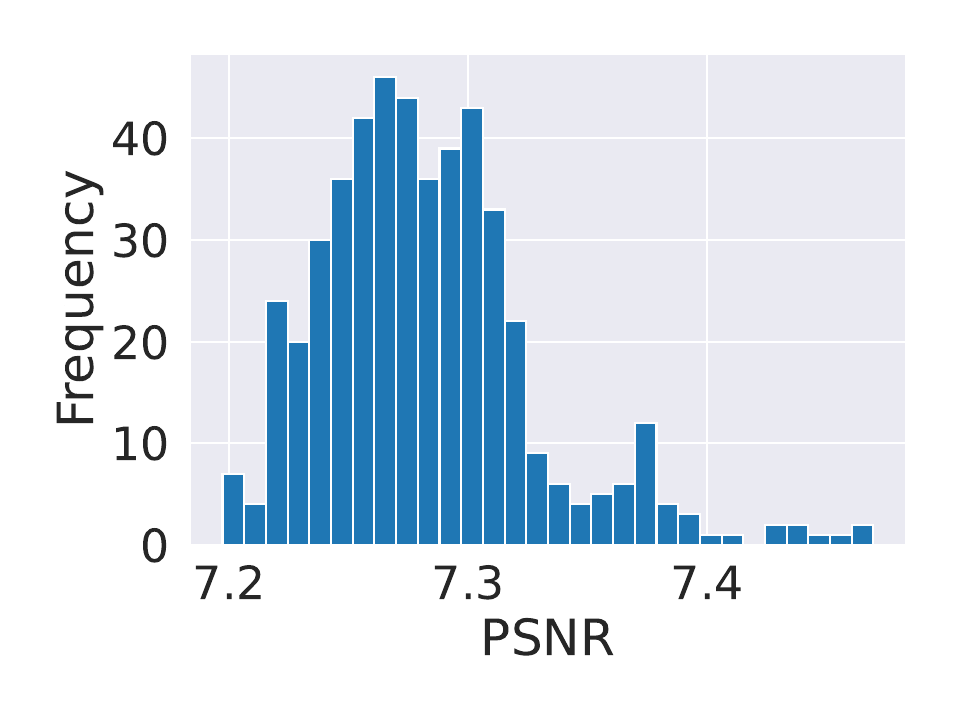}
    \caption{Backdoor PSNR}    
    \end{subfigure}
    \caption{Different histograms representing the distribution of SSIM and PSNR with clean and poisoned samples. The plots represent the robustness of the model to different prompts, meaning that there is no ``robust'' prompt that can improve the robustness of the model to compromised inputs. We would expect ``robust'' prompts to have PSNR or SSIM similar to the clean prompts under a backdoor input. However, we observe that all the combinations have smaller PSNR or SSIM than their clean counterparts.}
    \label{fig:prompt-engineering}
\end{figure*}

\section{Additional Results}
\label{app:additional_results}

\subsection{Injecting the Backdoor as a New Task}

Tables~\ref{tab:colorize-1},~\ref{tab:colorize-10}, and~\ref{tab:colorize-10-1} show the clean and backdoor performance of injecting the backdoor as a new task.

\subsection{Prompt Engineering}
Figure~\ref{fig:prompt-engineering} shows different results on the distribution of SSIM and PSNR for different prompt combinations for the prompt engineering defense.

\balance

\end{document}